\documentclass[aps,prl,reprint,superscriptaddress,longbibliography]{revtex4-2}

\usepackage[utf8]{inputenc}
\usepackage[T1]{fontenc}
\usepackage{amsmath,amssymb}
\usepackage{graphicx}
\usepackage{hyperref}
\usepackage{bm}
\usepackage{bbm}
\usepackage[dvipsnames]{xcolor}
\usepackage{physics}
\usepackage{placeins}
\usepackage{orcidlink}
\usepackage{array}
\usepackage{booktabs}
\usepackage{tabularx}

\newcommand{\vbm}[0]{${\rm V}_{\rm B}^-$ }
\newcommand{\gd}[0]{${\rm Gd}^{3+}$ }
\newcommand{\vv}[0]{VV$_0$ }
\newcommand{\sv}[0]{${\rm V}_{\rm Si}^-$ }

\makeatletter
\renewcommand{\p@subsection}{}
\renewcommand{\p@subsubsection}{}
\makeatother

\usepackage{tikz}
\usetikzlibrary{arrows.meta}

\tikzset{
  seqaxis/.style={line width=0.45pt},
  seqlabel/.style={font=\scriptsize, anchor=east},
  seqtext/.style={font=\scriptsize},
  laserinit/.style={
    draw=OliveGreen!70!black,
    fill=OliveGreen!30,
    rounded corners=0.6pt,
    minimum width=0.55cm,
    minimum height=0.20cm,
    inner sep=0pt
  },
  laserread/.style={
    draw=BrickRed!70!black,
    fill=BrickRed!30,
    rounded corners=0.6pt,
    minimum width=0.55cm,
    minimum height=0.20cm,
    inner sep=0pt
  },
  mwpulse/.style={
    draw=MidnightBlue!80!black,
    fill=MidnightBlue!35,
    minimum width=0.08cm,
    minimum height=0.34cm,
    inner sep=0pt
  },
  mwblock/.style={
    draw=MidnightBlue!80!black,
    fill=MidnightBlue!20,
    rounded corners=0.6pt,
    line width=0.45pt
  },
  seqarrow/.style={<->, line width=0.35pt}
}

\begin{document}

\title{Spin Relaxometry with Solid-State Defects: Theory, Platforms, and Applications}

\author{Ruotian Gong\,\orcidlink{0009-0006-6167-4326}}
\affiliation{Department of Physics, Washington University in St.~Louis, St.~Louis, MO 63130, USA.}

\author{Alex L.~Melendez\,\orcidlink{0009-0003-1610-1340}}
\email{melendezal@ornl.gov}
\thanks{Corresponding author.}
\affiliation{Center for Nanophase Materials Sciences, Oak Ridge National Laboratory, Oak Ridge, TN 37831, USA.}

\author{Guanghui He\,\orcidlink{0009-0009-7307-7694}}
\affiliation{Department of Physics, Washington University in St.~Louis, St.~Louis, MO 63130, USA.}
\affiliation{Center for Nanophase Materials Sciences, Oak Ridge National Laboratory, Oak Ridge, TN 37831, USA.}

\author{Zhongyuan Liu\,\orcidlink{0009-0000-4309-4527}}
\affiliation{Department of Physics, Washington University in St.~Louis, St.~Louis, MO 63130, USA.}

\author{Chong Zu\,\orcidlink{0000-0001-7803-1315}}
\email{zu@wustl.edu}
\thanks{Corresponding author.}
\affiliation{Department of Physics, Washington University in St.~Louis, St.~Louis, MO 63130, USA.}

\author{Huan Zhao\,\orcidlink{0000-0002-4982-0865}}
\email{zhaoh1@ornl.gov}
\thanks{Corresponding author.}
\affiliation{Center for Nanophase Materials Sciences, Oak Ridge National Laboratory, Oak Ridge, TN 37831, USA.}


%
\begin{abstract}
Spin relaxometry using solid-state spin defects, such as the diamond nitrogen-vacancy (NV) center, probes dynamical processes by measuring how environmental fluctuations enhance the spin relaxation rate. 
In the weak-coupling limit, relaxation rates sample the transverse magnetic noise power spectral density through a sensor-specific filter function, turning the defect into a local, frequency-selective noise spectrometer. 
This review bridges theory and experiment, clarifying how measured relaxation rates map onto noise spectra and how near-field geometry shapes the response. 
We highlight representative applications across condensed-matter physics, chemical and biological sensing, and relaxometry-based magnetic-resonance spectroscopy. We conclude with emerging opportunities and key challenges.
\end{abstract}


\maketitle


\section{Introduction}\label{sec:Introduction}

Many of the properties of material and biological systems are governed not only by static fields, but also by fluctuations. 
Thermally driven currents, fluctuating spins, molecular motion and more, all generate time-dependent magnetic fields whose spectra encode information from which physical properties can be extracted.
Relaxometry is the general strategy of sensing such fluctuations by the change in the rate at which a probe system returns to equilibrium. 
Rooted in the development of nuclear magnetic resonance in the 1940s, Bloch's phenomenological description of nuclear induction introduced the longitudinal and transverse relaxation times $T_1$ and $T_2$ as measurable quantities characterizing the equilibration of a spin ensemble
\cite{
    bloch1946nuclear,
    purcell1946resonance,
    rabi1938new,
    giunta2020discovery}.
Since then, relaxation rates have become widely used spectroscopic observables: the faster a spin relaxes, the stronger the environmental noise at frequencies capable of driving transitions.

The term relaxometry is often used most narrowly to refer to measurements of the longitudinal relaxation time $T_1$.
More broadly, however, spin systems can probe environmental fluctuations through several related decay channels.
Originally defined as the time constant over which the longitudinal (i.e., parallel to the quantization axis) magnetization component approaches its equilibrium value, $T_1$ is related to the component of environmental magnetic field noise that is transverse to the spin axis. 
It therefore describes the timescale for energy exchange/thermalization between the spin system and its environment, and is historically known as the spin-lattice relaxation time.
In contrast, the transverse relaxation time $T_2$ is referred to as the spin-spin relaxation, dephasing, or decoherence time, and similarly characterized the rate at which the transverse magnetization component decayed.
This dephasing is sensitive to fluctuations of the local transition frequency, most directly from magnetic-field noise parallel to the spin axis, as well as to static or slowly varying inhomogeneity in the case of the inhomogeneous dephasing time $T_2^\ast$.
Thus, relaxometry via $T_1$, $T_2$, $T_2^\ast$, and related rotating-frame measurements such as $T_{1\rho}$ provides complementary spectral windows into the same environmental fluctuations
\cite{
    degen2017quantum,
    barry2020sensitivity,
    mzyk2022relaxometry}.

The use of solid-state spin defects allows this idea to be extended to nanoscale sensing. 
Defects such as the diamond nitrogen-vacancy (NV) center have spin states that can be initialized, manipulated, and optically or electrically read out, allowing relaxation measurements to be performed without conventional inductive detection.
Due to the atomic-scale of the defect, its relaxation provides a highly spatially resolved probe of magnetic noise, enabling measurements of environmental dynamics near surfaces, interfaces, and heterogeneous nanoscale systems
\cite{
    doherty2013nitrogen,
    rondin2014magnetometry,
    degen2017quantum,
    barry2020sensitivity,
    casola2018probing}.

Three features make relaxometry with spin defects uniquely useful.
First, it is naturally a noise spectroscopy tool: in the weak-coupling regime, relaxation/dephasing rates are set by the environmental power spectral densities (PSDs) $S(\omega)$ at frequencies determined by the sensor transitions and applied control sequence.
Second, it is inherently local: the defect samples the near field of the environment, so the signal can be strongly distance dependent and spatially resolvable in scanning and wide-field geometries.
Lastly, optical polarization via an intersystem crossing coupled with fast non-invasive readout enables high throughput measurements allowing high sensitivity.

In practice, relaxometry is used in several complementary modes:
(i) \textbf{Local monitoring:} measure a relaxation time at a single defect, ensemble spot, or camera pixel to track changes in the local noise environment, such as variations in paramagnetic-ion concentration, or nearby spin density.
(ii) \textbf{Spatial imaging:} measure relaxation as a function of position to image spatial variations in noise, e.g., local defects, domain walls, current paths
\cite{
    steinert2013magnetic,
    tetienne2013spin,
    finco2021imaging,
    levine2019principles,
    mzyk2022relaxometry}.
(iii) \textbf{Spectroscopy via tuning:} sweep the static magnetic field, microwave drive, or other control parameter to tune the sensor transition or filter function and measure relaxation as a function of frequency, revealing resonant modes and cross-relaxation features 
\cite{
    wood2016wide,
    wood2017microwave,
    mignon2023fast}.

A key aspect is that relaxometry is typically more model-dependent than some other techniques, e.g., static-field sensing, wherein a measured resonance shift can be converted directly into a magnetic-field projection through a well-known gyromagnetic ratio.
In contrast, the relaxation rate must be connected to underlying material parameters through a complex and sample-specific model of (a) the relevant correlations in the sample (spins, currents, vortices, etc.) and (b) the geometry-dependent mapping from those sources to the field fluctuations at the defect location 
\cite{degen2017quantum,mzyk2022relaxometry}.
Such additional assumptions make the process of extracting physical quantities nontrivial.
Throughout this review we emphasize practical strategies---baseline subtraction, height dependence, field dependence, and independent knobs (temperature, drive power, device configuration)---that help make this inference robust.

Among the growing family of solid-state spin defects, the diamond NV center remains the most mature and widely used platform, and it therefore provides the central example throughout this review.
Its combination of room-temperature operation, optical initialization and readout, long spin lifetimes, and compatibility with shallow implantation or scanning-probe geometries has made it a benchmark system for nanoscale magnetic noise sensing
\cite{
doherty2013nitrogen,
rondin2014magnetometry,
degen2017quantum,
barry2020sensitivity}.
At the same time, many of the aforementioned concepts developed for NV relaxometry apply more broadly to other spin defects in diamond, hBN, SiC, and emerging host materials.

This review takes a ground-up approach to solid-state spin defect relaxometry, beginning with the experimental platforms used to implement spin-defect sensing.
These range from established NV sensing modalities to emerging host materials, including systems in which conventional optically detected magnetic resonance (ODMR) is weak or absent.
We next present a pedagogical framework that begins with the measured spin-dependent optical signal and builds toward the interpretation of relaxation and decoherence as noise-spectroscopy measurements, connecting pulse-sequence observables to sensor transition frequencies, magnetic noise power spectra, and ultimately to material response functions through geometry-dependent magnetic field propagation.
With this foundation in place, we survey representative applications of relaxometry across condensed-matter physics, spin spectroscopy, and biological environments. 
We conclude by identifying the central challenges and opportunities for the field, including the transition from qualitative contrast to quantitative inference, the control of surface and technical noise, improved throughput and calibration, operation in extreme or device-relevant environments, and the extension of relaxometry to new spin-defect platforms and geometries.

The central organizing principle is a simple translation: \emph{dynamics in a nearby system} $\rightarrow$ \emph{magnetic-field fluctuations at the sensor} $\rightarrow$ \emph{a change in the sensor's spin-relaxation time}. 
The key message is that relaxometry provides a local, spectrally selective probe of a sample's magnetic noise power spectral density (PSD) $S_B(\omega)$, containing information about microscopic dynamics and material parameters of the sample.
The conceptual map below summarizes the main concepts needed to make this translation quantitative.

\begin{itemize}

\item \textbf{Relaxometry measures field fluctuations, not static fields.}
In contrast to magnetometry, which measures static stray fields, relaxometry detects time-dependent magnetic noise produced by dynamical degrees of freedom.
The central object is therefore a magnetic noise PSD, rather than a static field map \cite{mzyk2022relaxometry,degen2017quantum}.

\item \textbf{Sensor converts magnetic noise into relaxation rate.}
E.g., in the weak-coupling regime, the contribution to the total longitudinal spin relaxation rate $\Gamma_1=1/T_1$ due to transverse magnetic noise from the sample is:
\begin{align}\label{eq:Gamma_to_S}
\Gamma_1^{\mathrm{sample}} \propto S_{B_\perp}^{\mathrm{sample}}(\omega_{\mathrm{s}}),
\end{align}
where $\omega_{\mathrm{s}}$ is the relevant spin-defect transition frequency.
This is the basic link between an excess decay rate and the local magnetic noise spectrum \cite{mzyk2022relaxometry,tetienne2013cavailles}.

\item \textbf{Spectral response of sensor determines noise component detected.}
Different measurement modes sample different frequency ranges of the PSD.
Roughly, $T_2^\ast$ probes near-DC and slowly varying noise, $T_2$ and dynamical-decoupling measurements probe kHz--MHz bands set by pulse spacing, $T_{1\rho}$ probes noise near the Rabi frequency, and $T_1$ probes transverse noise near the sensor transition frequency, typically in the MHz--GHz range depending on magnetic field 
\cite{
    degen2017quantum,
    barry2020sensitivity,
    mzyk2022relaxometry}.

\item \textbf{The sample--sensor geometry determines which fluctuations reach the defect.}
The sensor--sample distance, sensor orientation, and sample geometry act as spatial filters, so quantitative inference requires modeling the propagation of magnetic noise from the sample to the defect.
In near-field geometries, the sensor is especially sensitive to nearby sources and to spatial wavelengths comparable to or longer than the sensor--sample standoff distance
\cite{
    degen2017quantum,
    kolkowitz2015probing,
    mccullian2020broadband,
    takei2024detecting}.

\item \textbf{Material parameters are inferred through a source model.}
The measured relaxation rate constrains $S(\omega)$ at the sensor, but the desired physical quantities usually belong to the sample: spin correlation times, magnon spectra, conductivity, or other response functions.
Extracting such parameters requires a model that connects the sample's fluctuating degrees of freedom to the magnetic noise PSD measured by the defect.

\item \textbf{Baselines and control measurements determine interpretability.}
The measured longitudinal relaxation rate is typically a sum of multiple contributions,
\begin{align}\label{eq:total_relaxation_rate}
\Gamma_1^{\mathrm{meas}}
=
\Gamma_1^{\mathrm{bg}}
+
\Gamma_1^{\mathrm{sample}},
\end{align}
where intrinsic noise, surface-induced noise and other background contributions collectively captured by $\Gamma_1^{\mathrm{bg}}$ can all mimic or obscure the sample-induced relaxation $\Gamma_1^{\mathrm{sample}}$.
Control measurements and variation of relevant experimental parameters (e.g., distance, temperature, applied field, etc.) are therefore essential for isolating $\Gamma_1^{\mathrm{sample}}$ and assigning it to the physics of interest \cite{rosskopf2014investigation,romach2015spectroscopy,sangtawesin2019origins,cardoso2023impact}.

\end{itemize}
\vspace{0.5em}

The remainder of this review provides a practical and quantitative roadmap for implementing spin relaxometry, and surveys representative applications of the technique.

\begin{figure}[t]
    \centering
    \includegraphics[width=1\columnwidth]{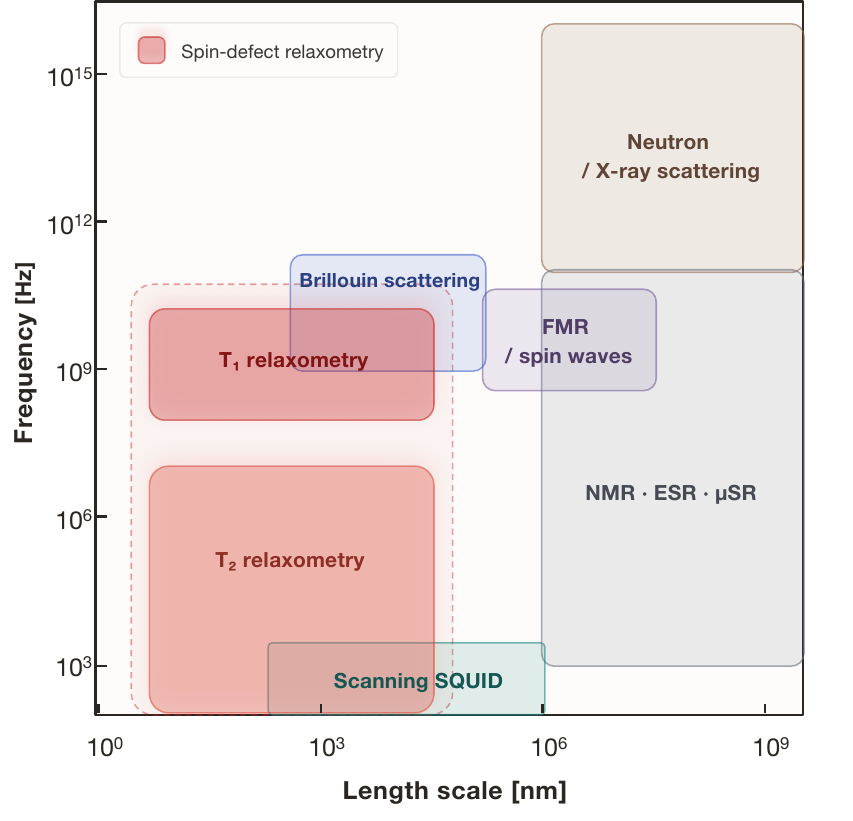}
    \caption{\textbf{Operating regimes of magnetic-noise sensing techniques in length-scale--fluctuation-frequency space.}
    Spin-defect relaxometry reaches nanoscale spatial resolution across a static-to-gigahertz window~\cite{degen2017quantum,casola2018probing,mzyk2022relaxometry}, overlapping the fluctuation phenomena targeted in this review while accessing length scales inaccessible to complementary probes: scanning SQUID~\cite{kirtley2010fundamental,vasyukov2013scanning}, Brillouin~\cite{sebastian2015microfocused} and FMR/spin-wave spectroscopy~\cite{maksymov2015broadband}, bulk NMR$\cdot$ESR$\cdot$$\mu$SR~\cite{kimmich2004field, hillier2022muon,schweiger2001principles}, and inelastic neutron/X-ray scattering~\cite{boothroyd2020principles,baron2020introduction}. The dashed outline marks the spin-defect relaxometry window.}
    \label{fig:landscape}
\end{figure}

\begin{figure*}[t]
    \centering
    \includegraphics[width=0.9\linewidth]{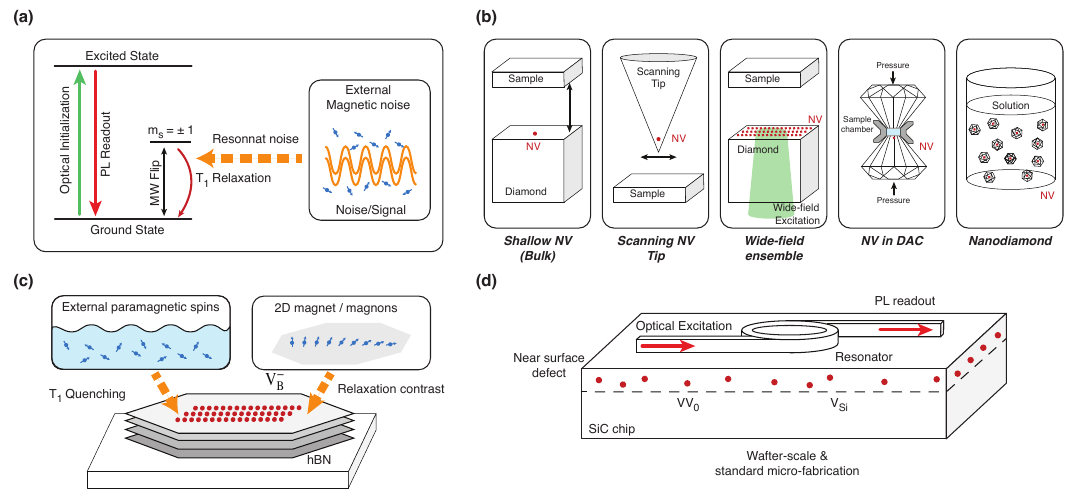}
    \caption{\textbf{Experimental platforms for spin relaxometry.}
    (a) Principle of spin relaxometry as frequency-selective noise detection: Optical pumping initializes the defect spin (e.g., NV) and photoluminescence readout monitors its population dynamics. Longitudinal relaxation ($T_1$) is enhanced by transverse magnetic noise with spectral weight resonant with the spin transition (tunable via bias field and, when used, microwave control), enabling spectroscopy of external targets through resonant energy exchange with the environment.
    (b) Modalities and hardware geometries of the diamond NV platform. All five configurations use NV centers, differing only in host geometry (from left to right) shallow near-surface NV centers in bulk diamond, scanning NV tips, wide-field NV ensembles, NV centers under high pressure in a diamond anvil cell (DAC), and fluorescent nanodiamonds in solution for relaxometric sensing in diverse environments.
    (c) Hexagonal boron nitride (hBN), as a two-dimensional spin-relaxometry platform, enables intrinsically small sensor--environment standoff distances, which is particularly advantageous in van der Waals heterostructures where hBN can be stacked directly with other 2D materials that host spin and charge noise.
    (d) Silicon carbide (SiC) is a promising host for chip-integrated spin relaxometry, leveraging wafer-scale processing, mature device fabrication, and defect spins that are compatible with on-chip photonic and electronic integration. Typical $T_1$/$T_2$ differ
    substantially across the three platforms (Table~\ref{tab:platforms}) and, for the NV, across the geometries in (b), most notably bulk vs.\ nanodiamond. (Sec.~\ref{sec:platform-nv})
    }
    \label{fig:platforms}
\end{figure*}

\section{Spin-defect relaxometry platforms}

Solid-state spin defects are point defects in crystals that possess a paramagnetic spin state and can be interrogated optically.
Crucially, many solid-state defects function at room temperature and can convert magnetic noise into optical fluorescence signals, enabling highly sensitive, non-invasive measurements with nanometer-scale resolution.
Below, we briefly survey the main solid-state spin-defect platforms used for relaxometry, namely diamond NV centers, spin defects in hexagonal boron nitride (hBN), defects in silicon carbide (SiC), and emerging new platforms, emphasizing what each enables experimentally and what limitations typically dominate. Table \ref{tab:platforms} summarizes defect creation types and typical spin parameters across representative relaxometry platforms.

We note that relaxation of a spin probe is well established for nuclear spins (NMR $T_1$/$T_2$ and field-cycling NMRD~\cite{kimmich2004field}), muons ($\mu$SR~\cite{hillier2022muon}), and electron spins (pulsed ESR~\cite{schweiger2001principles}), but these are bulk, large-volume, and sometimes high-field measurements. 
Defect-based relaxometry, which is the scope of this review, targets
the local magnetic-noise spectrum at the nanoscale. ~\cite{degen2017quantum,casola2018probing}.
It is complementary to these bulk techniques, trading quantitative robustness for nanoscale resolution, detection volumes, and room-temperature operation ~\cite{mzyk2022relaxometry,du2024single}.
Figure~\ref{fig:landscape} situates defect-based relaxometry among these complementary probes, mapping the accessible length scales and fluctuation frequencies and highlighting the nanoscale, room-temperature regime that motivates this review.

\subsection{Diamond NV Centers}\label{sec:platform-nv}

The NV center in diamond is the most established spin-defect platform for relaxometry and, more broadly, for nanoscale sensing.
It consists of a substitutional nitrogen atom adjacent to a carbon vacancy in the diamond lattice.
This defect has a spin-1 triplet ground state that can be polarized and read out optically at room temperature.
Being hosted in diamond's robust lattice, NV centers offer robust optical initialization and readout at room temperature, well-developed materials processing (implantation, delta-doping, surface termination), and long spin coherence and relaxation times \cite{doherty2013nitrogen, schirhagl2014nitrogen, barry2020sensitivity}.
In the context of relaxometry, NV centers can sensitively convert local fluctuating magnetic fields into changes in their fluorescence (Fig.~\ref{fig:platforms}a), allowing nanoscale sensing.
The NV based relaxometry was first demonstrated in 2013, when researchers showed that an NV's $T_1$ shortened in the presence of nearby paramagnetic ions, such as \gd \cite{steinert2013magnetic}.
Since then, NV relaxometry has been applied widely, and a practical advantage of diamond NVs is the diversity of sensor geometries (Fig.~\ref{fig:platforms}b):
\begin{itemize}
  \item \textbf{Single near-surface NVs in bulk diamond} maximize spatial resolution and coupling to local fluctuators, but can be limited by surface-related magnetic noise and charge instability \cite{rosskopf2014investigation,romach2015spectroscopy,sangtawesin2019origins}.
  \item \textbf{Scanning NV probes} enable controlled NV-sample distance and sub-diffraction-limit imaging, at the cost of slower acquisition and added drift/stability constraints \cite{maletinsky2012robust,rondin2014magnetometry}.
  \item \textbf{Wide-field NV ensembles} enable parallel relaxometry to be performed over large fields of view (including pixel-wise $T_1$ maps), with resolution set by diffraction, sensor depth, and pixel size \cite{levine2019principles,mzyk2022relaxometry}.
  \item \textbf{Diamond anvil cells (DACs)} enable quantum sensing under extreme pressures, allowing relaxometric probing of pressure-tuned magnetism and superconductivity in micron-scale samples \cite{hamlin2019extreme, hsieh2019imaging,bhattacharyya2024imaging, lesik2019magnetic, yip2019measuring}.
  \item \textbf{Nanodiamonds} provide straightforward integration into liquids and cells, but typically introduce larger inhomogeneity and more complex surface chemistry that can broaden $T_1$ distributions and add platform-specific backgrounds \cite{schirhagl2014nitrogen, perona2020nanodiamond}.
\end{itemize}

We note the NV coherence varies with host geometry: bulk, isotopically purified diamond yields the longest coherence times ($T_2$ up to ${\sim}1.8$\,ms at room temperature\cite{balasubramanian2009ultralong}).
Nearsurface NVs used for external sensing exhibit shortened coherence times due to surface-induced noise, ranging to tens-of-microseconds \cite{ohno2012engineering,rosskopf2014investigation,myers2014probing}.
Nanodiamonds show broadened, generally shorter $T_2$ distributions, typically on the order of microseconds arising from surface and size effects~\cite{tetienne2013spin,schirhagl2014nitrogen}. 
The bulk/optimized value quoted for NV in Table~\ref{tab:platforms} thus represents a best case---proximity- or integration-driven geometries (shallow, scanning, nanodiamond) trade coherence for resolution and deployability.

Despite its success, the NV platform has some limitations.
One challenge is the need to position NV centers close to the sample of interest (typically within $\sim$\,10\,nm of the diamond surface) to sense external spins. 
However, such shallow NVs often possess degraded coherence and relaxation times due to surface-induced noise, which can result in a loss of sensitivity \cite{rosskopf2014investigation,romach2015spectroscopy,sangtawesin2019origins}. 
Additionally, while diamond is an excellent host, its three-dimensionality means that integration with materials or devices can be non-trivial.
Nonetheless, the diamond NV center remains the gold standard for solid-state relaxometry, offering a proven combination of sensitivity, nanoscale resolution, and operation over a wide temperature and pressure range.

\subsection{hBN defects}

The aforementioned surface-noise and dimensionality challenges that often accompany NV centers motivate platforms where the spin is intrinsically at an interface.
Hexagonal boron nitride (hBN) is a wide band gap ($\sim$6 eV) van der Waals material that can be exfoliated to the few-layer limit and integrated into heterostructures, enabling ultra-small sensor-sample standoff without the need for near-surface defect engineering in a 3D crystal \cite{vaidya2023quantum}. 
The best-established hBN spin defect is the negatively charged boron vacancy \vbm, which exhibits room-temperature ODMR and coherent control \cite{gottscholl2020initialization, gottscholl2021room, gong2023coherent, gong2024isotope, liu2025temperature, biswas2025quantum}. 
The \vbm center consists of a missing boron atom in the lattice, where the negatively charged state results in a spin-1 system highly similar to NV center.
Beyond boron vacancies, other carbon-related defects in hBN have been reported to show ODMR as well (e.g., defects involving carbon substitution in the lattice) \cite{stern2022room,stern2024quantum}. 
While their structures are still being actively studied, these additional spin defects with brighter emission indicate that hBN may host a family of optically addressable spins with potentially diverse sensing properties \cite{chejanovsky2021single, mendelson2021identifying}.

Though still a young platform, hBN spin defects have already demonstrated the core capability for relaxometry.
A recent breakthrough showed that \vbm can detect external paramagnetic spins through a reduction of $T_1$ under ambient conditions \cite{gao2023quantum}.
Another exciting application of hBN defects is in probing magnetic phenomena in 2D heterostructures and devices (Fig.~\ref{fig:platforms}c).
\vbm centers have been used to sense spin wave (magnon) excitations in a magnetic layer through changes in the relaxation rate and resonance signals \cite{huang2022wide, zhou2024sensing}.
On the other hand, hBN spin defects face several challenges as an emerging platforms.
The intrinsic relaxation times of hBN defects are often over an order of magnitude shorter than those of NV. 
\vbm centers in particular are also much dimmer than NVs, which reduces sensitivity and necessitates the use of ensembles or longer averaging times for relaxometry measurements.
Despite these challenges, the rapid progress and demonstration of defects possessing certain advantages over NV (e.g., higher PL and PL contrast) position hBN as a flexible and promising 2D quantum sensing platform.

\newcolumntype{L}{>{\raggedright\arraybackslash}X} 
\newcolumntype{C}[1]{>{\hsize=#1\hsize\raggedright\arraybackslash}X}
\begin{table*}[t]
  \centering
  \footnotesize
  \setlength{\tabcolsep}{5pt}
  \renewcommand{\arraystretch}{1.3}
  \caption{Defect creation and typical spin parameters across representative relaxometry platforms. Values are room temperature unless noted. $T_2$ is defined with respect to the Hahn-echo signal and is extended further by dynamical decoupling (DD). Depth, density, and coherence are mutually constrained engineering choices rather than fixed platform constants---shallower or denser samples generally show shorter $T_2$.}
  \label{tab:platforms}
  \begin{tabularx}{\textwidth}{@{} l C{1.4} C{1.25} C{0.78} C{0.67} C{0.9} @{}}
    \toprule
    \textbf{Platform (defect)} & \textbf{Creation} & \textbf{Depth} & \textbf{Density} & \textbf{$T_1$} & \textbf{$T_2$} \\
    \midrule
    Diamond: NV$^-$
      & Native growth, implantation, irradiation, CVD $\delta$-doping~\cite{pezzagna2010creation,ohno2012engineering,barry2020sensitivity}
      & ${<}\,10\,$nm--$10\,\mu$m ($\delta$-doping 5-100\,nm)~\cite{pezzagna2010creation,ohno2012engineering}
      & single $\to\,{\sim}\,10^{18}\,$cm$^{-3}$~\cite{barry2020sensitivity}
      & \textbf{${\sim}$\,ms} ($\to$\,min, cryo)~\cite{jarmola2012temperature}
      & \textbf{${\sim}\,\vb*{\mu}$s--ms} (1.8\,ms iso.)~\cite{balasubramanian2009ultralong} \\
    \addlinespace
    hBN: \vbm
    & Neutron / electron / ion irradiation, fs-laser~\cite{gottscholl2020initialization,gottscholl2021room,guo2022generation}
          & nm-scale (${\approx}$ flake thickness)~\cite{guo2022generation}
          & ensemble
          & \textbf{${\sim}\,18\,\vb*{\mu}$s}~\cite{gottscholl2021room,liu2025temperature}
          & \textbf{${\sim}\,50$\,ns} ($\to$\,tens\,$\mu$s, DD)~\cite{gottscholl2021room,rizzato2023extending, gong2024isotope} \\
    \addlinespace
    SiC: VV$_0$, \sv
      & Irradiation or implantation $+$ anneal, CVD~\cite{koehl2011room,christle2015isolated,son2020developing}
      & ${\sim}$\,nm to several $\mu$m~\cite{son2020developing}
      & single $\to$ ensemble~\cite{christle2015isolated,widmann2015coherent}
      & \textbf{ms--s}~\cite{son2020developing}
      & \textbf{${\sim}\,1\,$ms} ($>$\,5\,s, DD)~\cite{christle2015isolated,anderson2022five} \\
    \bottomrule
  \end{tabularx}
\end{table*}

\subsection{SiC and Other Emerging Defects}

Like diamond and hBN, silicon carbide (SiC) is a wide-bandgap semiconductor that hosts several important spin defects.
Two well-known spin-active centers in SiC are the neutral divacancy (VSiV$^0_{\rm C}$, often abbreviated as \vv) and the silicon vacancy (\sv) \cite{widmann2015coherent, wolfowicz2017optical, nagy2019high, christle2017isolated}.
The divacancy in 4H-SiC consists of an adjacent carbon vacancy and silicon vacancy pair, and is isoelectronic to the NV and \vbm (i.e., a spin-1 triplet with optical transitions that enable spin polarization and readout).
The negatively charged silicon vacancy \sv (often at the so-called V2 site in 4H-SiC) is a spin-3/2 system that also exhibits ODMR at room temperature.
A major appeal of SiC as a host is the availability of high-quality, wafer-scale material and established microfabrication techniques.
In other words, SiC defects can be integrated into devices (e.g. nanophotonic cavities, electrical circuits) using standard semiconductor processing (Fig.~\ref{fig:platforms}d).
Both \vv and \sv centers have demonstrated long spin coherence times (up to milliseconds) in isotope-purified SiC and can be engineered at specific sites or depths via ion implantation \cite{christle2017isolated, seo2016quantum}.
Although research on SiC defects initially focused on quantum communication and computing, there is growing interest in their use for relaxometry-based sensing.
Recent work has pushed SiC divacancy qubits toward practical relaxometry: near-surface defects have been used for $T_1$-based detection of surface paramagnetic species, highlighting SiC as a promising, device-compatible alternative for relaxometry \cite{li2025non}.
SiC spin defects offer clear advantages like compatibility with existing tech and bio-optical windows, but requires continued improvements in defect engineering and surface science to fully match the sensitivity of conventional counterparts.

Beyond NV, hBN, and SiC, a broader landscape of optically addressable spins is under active development. 
However, most remain at the level of sensing demonstrations rather than established, quantitative nanoscale relaxometry.
For example, rare-earth-ion-doped crystals (e.g., Eu/Pr/Er in oxide hosts) offer exceptionally narrow optical transitions and long-lived spin states at cryogenic temperatures, and are widely used for quantum memories and coherent spectroscopy \cite{thiel2011rare,bertaina2007rare, siyushev2014coherent}.
However, most sensing demonstrations to date rely on spectral shifts / coherent optical spectroscopy rather than relaxometry, and the need for cryogenic operation remains a major practical constraint.
Another prospective 2D host is the family of transition-metal dichalcogenides (TMDs), e.g., WS$_2$, WSe$_2$, MoTe$_2$, where extensive defect engineering and single-photon-emitter (SPE) work already exists, including SPEs operating at technologically relevant telecom wavelengths \cite{zhao2025telecom,zhao2021site, chen2025low}.
Several first-principles studies have specifically proposed spin-triplet defect candidates in monolayer TMDs (e.g., antisite defects, or transition-metal substitutions at chalcogen sites) with optical transitions that could be compatible with spin readout \cite{tsai2022antisite,shang2022first, lee2022spin}.
At present, however, these remain largely computational blueprints, and robust room-temperature ODMR and quantitative $T_1$-relaxometry demonstrations in TMD defects are still much less established.

\subsection{Typical numbers: sensitivity, accuracy, and readout mode}

As a magnetometer, a single NV reaches $\sim$\,$\mu$T/$\sqrt{\text{Hz}}$ (DC) to tens of nT/$\sqrt{\text{Hz}}$ (AC), while dense ensembles ($\sim\!10^{11}$--$10^{12}$ defects) approach the sub-pT/$\sqrt{\text{Hz}}$ regime \cite{barry2020sensitivity}.
For relaxometry, a single shallow or nanodiamond NV samples a volume of order $(5$--$10\,\text{nm})^3$ and resolves $\sim$\,14 external electron spins in $\sim$\,10\,s (single-electron-spin noise sensing demonstrated with a single NV in a $10$\,nm nanodiamond) \cite{tetienne2013spin}, while a near-surface NV ensemble images $\sim$\,$10^3$ statistically polarized spins per diffraction-limited ($\lesssim$\,500\,nm) pixel \cite{steinert2013magnetic}.
In solution, nanodiamond $T_1$-relaxometry detects paramagnetic ion (e.g.~Gd$^{3+}$) concentrations across the nanomolar-to-millimolar range, with $T_1$ dropping from hundreds of $\mu$s in pure water to tens of $\mu$s at high concentration~\cite{schirhagl2014nitrogen,mzyk2022relaxometry}.
The relaxation time itself is extracted from stretched exponential fits to a typical accuracy of $\sim$\,5--15\%, limited by photon shot noise and the number of averages. 
The dominant systematic uncertainty is usually background/baseline subtraction (Eq.~\eqref{eq:total_relaxation_rate}) and the source model, not the fit.
Finally, because optical readout yields $\ll$\,1 spin-projection photon per shot, single-NV relaxometry is an inherently repeated, averaged measurement ($\sim$\,$10^4$--$10^7$ repetitions, seconds to minutes per point) that maximizes ($\sim$\,nm) spatial resolution, whereas wide-field/ensemble configurations read out many defects in parallel for per-pixel $T_1$ maps at diffraction-limited resolution and higher throughput. 
As previously mentioned, $V_B^-$ in hBN are dim with low ODMR contrast ($<$\,5\%) and is used almost exclusively in ensembles, while SiC spans single-defect to ensemble operation \cite{gottscholl2021room,christle2015isolated}.
Table~\ref{tab:figures_of_merit} collects these application-level figures of merit across configurations.

\newcolumntype{Z}[1]{>{\hsize=#1\hsize\raggedright\arraybackslash}X}

\begin{table*}[t]
  \centering
  \footnotesize
  \setlength{\tabcolsep}{4pt}
  \renewcommand{\arraystretch}{1.3}
  \caption{Representative relaxometry figures of merit by sensor configuration, complementing the intrinsic spin parameters of Table~\ref{tab:platforms}. Values are demonstrated examples from the cited works, not platform constants. Sensitivity, detection limit, and accuracy depend strongly on standoff, defect density, readout scheme, and averaging time. Field sensitivities are magnetometry references included for context. Entries marked ``demo.'' indicate single proof-of-principle demonstrations. hBN/SiC relaxometry figures of merit remain far less characterized than for NV.}
  \label{tab:figures_of_merit}
  \begin{tabularx}{\textwidth}{@{} l Z{1.05} Z{0.9} Z{1.25} Z{0.95} Z{0.85} @{}}
    \toprule
    \textbf{Configuration} & \textbf{Readout \& averaging} & \textbf{Resolution / standoff} & \textbf{Detection limit (demonstrated)} & \textbf{Field sensitivity} & \textbf{$T_1$ accuracy} \\
    \midrule
    Single NV
      (near-surface / nanodiamond)
      & Repeated single-spin readout; $\sim\!10^4$--$10^7$ averages; s--min per point
      & $\sim\!5$--$10\,$nm standoff; $\sim\!(5$--$10\,$nm$)^3$ volume
      & $\sim\!14$ electron spins in $10\,$s; single external e$^-$ spin~\cite{tetienne2013spin}
      & $\sim\!\mu$T\,Hz$^{-1/2}$ (DC); tens\,nT\,Hz$^{-1/2}$ (AC)~\cite{barry2020sensitivity}
      & $\sim\!5$--$15\%$ \\
    \addlinespace
    NV ensemble / wide-field
      & Parallel camera readout; per-pixel $T_1$ maps; density $\to\!\sim\!10^{18}$\,cm$^{-3}$
      & Diffraction-limited, $\lesssim\!500\,$nm
      & $\sim\!10^3$ statistically polarized spins/pixel ($\sim\!32$ net)~\cite{steinert2013magnetic}
      & pT--sub-pT\,Hz$^{-1/2}$~\cite{barry2020sensitivity}
      & $\sim\!5$--$15\%$ \\
    \addlinespace
    hBN: $V_B^-$
      & Ensemble only; low contrast ($<\!5\%$); short $T_1\!\sim\!18\,\mu$s
      & $\sim\!$flake thickness (nm)
      & Paramagnetic ions in liquid (demo.)~\cite{gao2023quantum}
      & $\sim\!10$--$100\,\mu$T\,Hz$^{-1/2}$~\cite{gottscholl2021room}
      & not yet established \\
    \addlinespace
    SiC: VV$_0$, V$_{\rm Si}$
      & Single $\to$ ensemble; long $T_1$ (ms--s)
      & $\sim\!$nm--$\mu$m depth
      & Surface paramagnetic species via $T_1$ (demo.)~\cite{li2025non}
      & $\sim\!\mu$T\,Hz$^{-1/2}$ (emerging)~\cite{li2025non}
      & not yet established \\
    \bottomrule
  \end{tabularx}
\end{table*}


\begin{figure*}[t]
    \centering
    \includegraphics[width=0.95\linewidth]{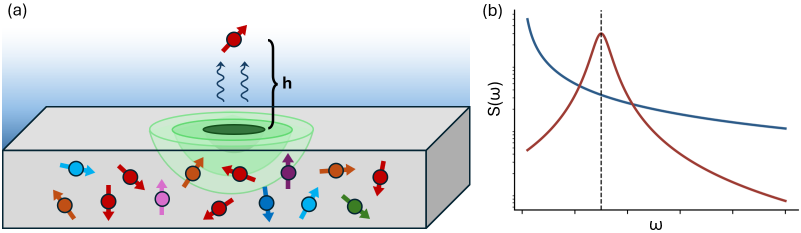}
    \caption{
    (a) Schematic of a single single spin sensor above a sample containing diverse fluctuating spins. Local magnetic noise spectrum is depicted by blue gradient/squiggly arrows. The sensor spin is sensitive to a sample volume illustrated by the concentric green hemispheres, with stronger coupling to spins near the center (dark green). The distance $h$ sets the spatial resolution of the sensor.
    (b) Schematic of overlap between power spectral density of sample (blue) with that of the spin defect (red). Center (ODMR) frequency of spin defect is represented by the vertical gray dashed line.
    }
    \label{fig:theory}
\end{figure*}

\section{Theory of Spin Relaxometry}\label{sec:Theory}

In this section, we outline both the experimental protocols and the basic theoretical framework of spin relaxometry with solid-state defects.
We use the term ``spin relaxometry'' in a broad sense to include both longitudinal population relaxation and coherence-decay measurements.
Thus, $T_1$ refers specifically to longitudinal relaxation, while $T_2^\ast$, $T_2$, and $T_{1\rho}$ describe transverse or rotating-frame coherence decay.
These quantities are physically distinct, but they are closely related as noise-spectroscopy tools: each converts environmental fluctuations into a measurable change in the defect spin state.

The section is organized as follows.
We first introduce the basic all-optical $T_1$ measurement, emphasizing optical initialization, spin-dependent readout, and the extraction of relaxation rates from observable PL decay curves.
We then describe microwave-controlled protocols, including Rabi, Ramsey, echo, inversion-recovery, and spin-lock measurements, which prepare and probe different spin populations or coherences.
Next, we introduce the NV spin Hamiltonian and show how the defect transition frequencies determine which components and frequencies of environmental noise are sampled.
We then connect $T_1$ relaxation to the transverse magnetic noise power spectral density at the defect location, discuss how different control sequences provide complementary spectral windows, and finally outline how local magnetic noise is related to microscopic material response functions through geometry-dependent magnetic field propagators.

\subsection{Basic Relaxometry}

To describe relaxation, we must first consider the ground state of a spin defect such as the NV center and the populations of its spin states at equilibrium, then how a non-equilibrium state is prepared and subsequently relaxes.
The triplet electronic ground state of the NV$^-$ consists of three spin states $m_s=0,\pm1$.
In zero applied field the spin quantization axis lies along the N--V bond, with the $m_s=\pm1$ states lying at an energy of $h\times(2.87\,\text{GHz})\approx k_B\times(0.14\,\text{K})$ above the $m_s=0$ state.
Here $h$ is Planck's constant and $k_B$ is Boltzmann's constant.
Hence $D=2\pi\times2.87\,$GHz is referred to as the zero-field splitting, and under experimental conditions well above this temperature all three spin states are approximately equally populated in thermal equilibrium.

Thus the simplest form of relaxometry, known as an ``all-optical'' (i.e., without MW) $T_1$ measurement, is given by the basic sequence:
\begin{equation}\label{eq:all_optical_T1}
\vcenter{\hbox{
\begin{tikzpicture}[
    x=0.88cm,
    y=0.60cm,
    baseline=(current bounding box.center),
    font=\scriptsize
]
\draw[seqaxis] (0,0) -- (5.2,0);
\draw[seqaxis] (0,-0.75) -- (5.2,-0.75);
\node[seqlabel] at (-0.15,0) {Laser};
\node[seqlabel] at (-0.15,-0.75) {MW};

\node[laserinit,label=above:{init.}] at (0.75,0) {};
\node[laserread,label=above:{r.o.}] at (4.45,0) {};

\draw[seqarrow] (1.15,0.30) -- node[above] {$\tau$} (4.05,0.30);
\end{tikzpicture}
}}
\end{equation}
The experimental observable is the change in the PL as $\tau$ is varied, and is typically cast as the photoluminescence (PL) contrast $C(\tau)=\mathrm{PL}(\tau)/\mathrm{PL}(0)$.
Here the same green wavelength is used in both the initialization and readout (r.o.) laser pulses, though the readout is drawn as red to emphasize that the defect PL is being measured.
In this and other sequences \eqref{eq:rabi_sequence}--\eqref{eq:T1rho_sequence}, pulse lengths are not drawn to scale.

The initialization pulse of \eqref{eq:all_optical_T1} prepares the defect in a non-equilibrium spin population state.
For the NV center, optical excitation preferentially polarizes the ground-state spin into the $m_s=0$ state through a spin-selective intersystem crossing.
Bottlenecked by the lifetime of the metastable singlet state ($\sim$\,250\,ns), optical initialization requires $\sim$\,1\,$\mu$s of laser illumination.
Given that the absorption and emission phonon sidebands of the NV are peaked at roughly 570\,nm and 690\,nm respectively, optical excitation is usually done with a green laser which can later be easily filtered out to collect only the red PL spectrum.
After initialization, the laser is turned off and the system is allowed to relax back to equilibrium for a time $\tau$.
The PL contrast depends on $\tau$ because a higher population of $m_s=0$ produces a higher PL for NVs during readout.
Thus as the $m_s=0$ population returns to equilibrium ($p_0\approx1/3$), a PL decay curve $C(\tau)$ is observed.

Optical readout is therefore both a measurement and a perturbation.
Because the excited-state lifetime is only $\sim$\,10--13\,ns, fluorescence is effectively emitted only while the laser is applied.
However, the same laser pulse that produces PL also drives the spin-selective intersystem crossing and begins to repolarize the NV into $m_s=0$.
As a result, the PL is approximately proportional to the pre-readout spin population only during a short window before substantial optical repolarization begins.
Only a limited number of useful photons can therefore be collected in a single shot, i.e., before the sequence must be repeated.
Photon shot noise, together with the long wait times required when $T_1$ lies in the $\mu$s--ms range (during which no photons are collected), can make all-optical $T_1$ measurements slow, especially for spatial imaging, field sweeps, or measurements requiring high signal-to-noise ratio
\cite{
    doherty2013nitrogen,
    rondin2014magnetometry,
    barry2020sensitivity}.

Relaxation can often be modeled by a single effective transition rate, obeying a first-order rate equation:
\begin{align}\label{eq:rate_equation}
\frac{d p_0}{d\tau}
=
-\frac{p_0(\tau)-p_{0}^{\mathrm{eq}}}{T_1},
\end{align}
where $p_0(\tau)$ is the population in the optically bright $m_s=0$ state and $p_{0}^{\mathrm{eq}}$ is its long-time equilibrium value.
Assuming a sufficiently short readout window such that the measured PL is approximately proportional to the spin population $p_0(\tau)\propto C(\tau)$, the solution of \eqref{eq:rate_equation} describes the PL contrast as a single exponential decay with time constant $T_1$. 
However in practice, shallow defects, nanodiamonds, and ensembles often sample a distribution of local environments or multiple relaxation channels, in which case stretched-exponential or multi-exponential forms may be more appropriate.
The measured contrast is therefore often fit phenomenologically to a stretched-exponential form
\begin{align}\label{eq:PL_decay_curve}
C(\tau) = C_\infty + Ae^{-(\tau/T_1)^\beta},
\end{align}
where $C_\infty$ is the equilibrium value, $A=(C_0-C_\infty)$ is the amplitude with $C_0$ being the initial PL contrast, $T_1=1/\Gamma_1$ is the longitudinal relaxation time and $\beta$ is a phenomenological stretching exponent.
Here $\beta<1$ often indicates a broad distribution of relaxation rates or local environments, while $\beta>1$ can arise from correlated or non-Markovian dynamics.
Thus, while the single-exponential model provides the basic starting point for interpreting $T_1$ relaxometry, stretched-exponential or multi-exponential forms are often more apt models for real samples
\cite{
    mzyk2022relaxometry,
    tetienne2013cavailles}.

A crucial point is that the measured rate generally contains several contributions.
The total relaxation rate is a sum of rates of the form of Eq.~\eqref{eq:total_relaxation_rate}.
In practice, $\Gamma_1^{\mathrm{sample}}$ is usually inferred from an excess relaxation rate, for example by comparing measurements on and off the sample, before and after adding a target species, or as a function of sensor--sample distance, temperature, magnetic field, or device state.
The task of theory is therefore twofold: first, to relate the sample-induced rate to the magnetic noise spectrum at the defect location, and second, to connect that local magnetic noise to the microscopic degrees of freedom in the sample (shown schematically in Fig.~\ref{fig:theory}(a)).

Although the discussion above used the NV center as the primary example, the basic logic of relaxometry is more general.
Any solid-state spin defect can, in principle, be used for relaxometry if it provides three ingredients: preparation of a non-equilibrium spin population or coherence, evolution under environmental noise, and a readout channel sensitive to the final spin state.
The microscopic mechanisms that provide these ingredients can differ substantially between platforms.
For example, the optical initialization and spin-dependent PL readout of the NV center rely on its spin-selective intersystem crossing, whereas other defects may require different optical transitions, resonant excitation, electrical readout, cryogenic operation, or may offer weaker spin contrast.
Thus, the specific pulse sequences, wavelengths, contrast levels, and accessible frequency ranges are platform-dependent, but the central principle remains the same: relaxation or decoherence of the defect spin converts environmental noise into a measurable change in the readout signal.

\subsection{Microwave-Controlled Relaxometry and Coherence Measurements}

The all-optical $T_1$ sequence described above measures the relaxation of an optically initialized spin population without applying resonant microwave control.
However, microwave pulses are crucial for preparing specific spin states, rotating coherent superpositions, or selectively probing individual transitions.
For the NV center, this is typically done by applying a microwave field resonant with either the $m_s=0\leftrightarrow+1$ or $m_s=0\leftrightarrow-1$ transition.
Such microwave-controlled sequences are useful both for calibrating the sensor and for accessing additional relaxation and dephasing timescales
\cite{
    degen2017quantum,
    barry2020sensitivity,
    mzyk2022relaxometry}.

The basic calibration experiment is a Rabi measurement.
After optical initialization into $m_s=0$, a resonant microwave pulse of variable duration $t$ is applied, followed by optical readout:
\begin{equation}\label{eq:rabi_sequence}
\vcenter{\hbox{
\begin{tikzpicture}[
    x=0.88cm,
    y=0.60cm,
    baseline=(current bounding box.center),
    font=\scriptsize
]
\draw[seqaxis] (0,0) -- (5.2,0);
\draw[seqaxis] (0,-0.75) -- (5.2,-0.75);
\node[seqlabel] at (-0.15,0) {Laser};
\node[seqlabel] at (-0.15,-0.75) {MW};

\node[laserinit,label=above:{init.}] at (0.75,0) {};
\node[laserread,label=above:{r.o.}] at (4.45,0) {};

\draw[mwblock] (1.15,-1.00) rectangle (4.05,-0.50);
\node[seqtext] at (2.45,-0.75) {$\omega_{\mathrm{NV}}$};
\draw[seqarrow] (1.15,-1.25) -- node[below] {$t$} (4.05,-1.25);
\end{tikzpicture}
}}
\end{equation}
Here $\omega_\text{NV}$ represents either ODMR frequency $\omega_{0\leftrightarrow+1}$ or $\omega_{0\leftrightarrow-1}$. 
As $t$ is varied, the spin population oscillates between $m_s=0$ and the driven $m_s=\pm1$ state.
The corresponding PL contrast is described by a decaying oscillatory signal:
\begin{align}\label{eq:rabi_signal}
C(t) = C_\infty + Ae^{-\left(t/T_R\right)^{\beta}}
\cos\!\left(\Omega_R t+\phi\right),
\end{align}
where $\Omega_R$ is the Rabi frequency, $T_R$ describes the decay of the driven oscillation envelope, and $\phi$ accounts for pulse-phase and timing offsets.
The Rabi frequency is proportional to the strength of the microwave field transverse to the NV axis in the rotating-wave approximation (RWA).
The time required to rotate the spin population completely from one state to the other defines a $\pi$ pulse, while half of this duration defines a $\pi/2$ pulse, so-called due to the angle rotated on the Bloch sphere.
Determination of these durations is crucial for any pulse sequence involving controlled rotations, and deviations from them will directly impact the state fidelity.
The decay of the Rabi envelope reflects the lifetime of driven coherent oscillations and can be limited by detuning, drive noise, and intrinsic relaxation/dephasing. 
It is therefore related to, but not identical to, the free-induction dephasing time $T_2^\ast$, which is conventionally measured with a Ramsey sequence.

A Ramsey measurement uses two $\pi/2$ pulses separated by a free-evolution time $\tau$:
\begin{equation}\label{eq:ramsey_sequence}
\vcenter{\hbox{
\begin{tikzpicture}[
    x=0.86cm,
    y=0.60cm,
    baseline=(current bounding box.center),
    font=\scriptsize
]
\draw[seqaxis] (0,0) -- (5.4,0);
\draw[seqaxis] (0,-0.75) -- (5.4,-0.75);
\node[seqlabel] at (-0.15,0) {Laser};
\node[seqlabel] at (-0.15,-0.75) {MW};

\node[laserinit,label=above:{init.}] at (0.75,0) {};
\node[laserread,label=above:{r.o.}] at (4.65,0) {};

\draw[mwblock] (1.15,-1.00) rectangle (1.80,-0.50);
\node[seqtext] at (1.475,-0.75) {$\pi/2$};

\draw[mwblock] (3.60,-1.00) rectangle (4.25,-0.50);
\node[seqtext] at (3.925,-0.75) {$\pi/2$};

\draw[seqarrow] (1.80,-1.25) -- node[below] {$\tau$} (3.60,-1.25);
\end{tikzpicture}
}}
\end{equation}
The first $\pi/2$ pulse creates a coherent superposition of the two driven spin states.
During the free-evolution time, this superposition accumulates phase relative to the microwave reference frame at a rate set by the detuning between the microwave frequency and the local spin transition frequency.
The second $\pi/2$ pulse converts the accumulated phase back into a population difference that can be read out optically.
The resulting Ramsey signal can be fit to an equation of the form \eqref{eq:rabi_signal} with $\{T_R,\,\Omega_R\}$ replaced by $\{T_2^\ast,\,\Delta\}$, where $\Delta=\omega-\omega_\text{NV}$ is the detuning between the microwave drive $\omega$ and the spin transition frequency $\omega_\text{NV}$, and $T_2^\ast$ is the inhomogeneous dephasing time.
Spatially varying or slowly fluctuating transition frequencies cause different experimental repetitions or different defects in an ensemble to acquire different phases, leading to a decay of the Ramsey contrast.

Microwave pulses can also be used in $T_1$ relaxometry.
A common microwave-assisted version is an inversion-recovery measurement, in which a $\pi$ pulse first transfers population from the optically bright $m_s=0$ state into one of the darker $\pm1$ states:
\begin{equation}\label{eq:mw_T1_sequence}
\vcenter{\hbox{
\begin{tikzpicture}[
    x=0.86cm,
    y=0.60cm,
    baseline=(current bounding box.center),
    font=\scriptsize
]
\draw[seqaxis] (0,0) -- (5.4,0);
\draw[seqaxis] (0,-0.75) -- (5.4,-0.75);
\node[seqlabel] at (-0.15,0) {Laser};
\node[seqlabel] at (-0.15,-0.75) {MW};

\node[laserinit,label=above:{init.}] at (0.75,0) {};
\node[laserread,label=above:{r.o.}] at (4.65,0) {};

\draw[mwblock] (1.15,-1.00) rectangle (1.65,-0.50);
\node[seqtext] at (1.40,-0.75) {$\pi$};

\draw[seqarrow] (1.75,0.30) -- node[above] {$\tau$} (4.20,0.30);
\end{tikzpicture}
}}
\end{equation}
During the waiting time, longitudinal relaxation drives the spin population back toward equilibrium.
The resulting recovery of the PL as a function of $\tau$ provides another measurement of $T_1$ and is described by Eq.~\eqref{eq:PL_decay_curve}.
Compared with an all-optical $T_1$ measurement, microwave-assisted protocols can prepare a more specific initial spin population and can selectively address one of the $\ket{0}\leftrightarrow\ket{\pm1}$ transitions.
For this reason, they are useful when one wants to distinguish relaxation channels associated with different transitions, although they require resonant microwave control and can be sensitive to microwave heating or pulse imperfections.

The Hahn-echo sequence is the simplest microwave protocol for reducing sensitivity to static or slowly varying frequency shifts.
It consists of a $\pi/2$ pulse, a free-evolution period, a $\pi$ pulse, a second free-evolution period, and a final $\pi/2$ pulse:
\begin{equation}\label{eq:spin_echo_sequence}
\vcenter{\hbox{
\begin{tikzpicture}[
    x=0.78cm,
    y=0.60cm,
    baseline=(current bounding box.center),
    font=\scriptsize
]
\draw[seqaxis] (0,0) -- (6.8,0);
\draw[seqaxis] (0,-0.75) -- (6.8,-0.75);
\node[seqlabel] at (-0.15,0) {Laser};
\node[seqlabel] at (-0.15,-0.75) {MW};

\node[laserinit,label=above:{init.}] at (0.75,0) {};
\node[laserread,label=above:{r.o.}] at (6.00,0) {};

\draw[mwblock] (1.15,-1.00) rectangle (1.80,-0.50);
\node[seqtext] at (1.475,-0.75) {$\pi/2$};

\draw[mwblock] (3.075,-1.00) rectangle (3.725,-0.50);
\node[seqtext] at (3.40,-0.75) {$\pi$};

\draw[mwblock] (5.00,-1.00) rectangle (5.65,-0.50);
\node[seqtext] at (5.325,-0.75) {$\pi/2$};

\draw[seqarrow] (1.80,-1.25) -- node[below] {$\tau/2$} (3.075,-1.25);
\draw[seqarrow] (3.725,-1.25) -- node[below] {$\tau/2$} (5.00,-1.25);
\end{tikzpicture}
}}
\end{equation}
The central $\pi$ pulse reverses the phase accumulation caused by static detuning, so that quasi-static inhomogeneity is largely refocused at the end of the sequence.
The decay of the echo contrast is fit to an equation of the form \eqref{eq:PL_decay_curve}, except in this case the decay constant represents $T_2^\mathrm{echo}$ rather than $T_1$.
Often called just $T_2$, it is more precisely the Hahn-echo coherence time.
The decay reflects noise that changes during the sequence, as well as pulse errors and other nonidealities.
More generally, the measured coherence time depends on the control sequence: multi-pulse dynamical-decoupling protocols can extend the apparent coherence time and shift the measurement sensitivity to different noise frequencies.

A rotating-frame relaxation measurement, denoted $T_{1\rho}$, uses a continuous resonant microwave drive to ``lock'' the spin along an axis in the rotating frame.
A typical sequence is
\begin{equation}\label{eq:T1rho_sequence}
\vcenter{\hbox{
\begin{tikzpicture}[
    x=0.78cm,
    y=0.60cm,
    baseline=(current bounding box.center),
    font=\scriptsize
]
\draw[seqaxis] (0,0) -- (6.8,0);
\draw[seqaxis] (0,-0.75) -- (6.8,-0.75);
\node[seqlabel] at (-0.15,0) {Laser};
\node[seqlabel] at (-0.15,-0.75) {MW};

\node[laserinit,label=above:{init.}] at (0.75,0) {};
\node[laserread,label=above:{r.o.}] at (6.10,0) {};

\draw[mwblock] (1.15,-1.00) rectangle (1.80,-0.50);
\node[seqtext] at (1.475,-0.75) {$\pi/2$};

\draw[mwblock] (1.95,-1.00) rectangle (4.90,-0.50);
\node[seqtext] at (3.425,-0.75) {spin lock};

\draw[mwblock] (5.05,-1.00) rectangle (5.70,-0.50);
\node[seqtext] at (5.375,-0.75) {$\pi/2$};

\draw[seqarrow] (1.95,-1.25) -- node[below] {$\tau$} (4.90,-1.25);
\end{tikzpicture}
}}
\end{equation}
Here the spin-lock drive is a continuous resonant microwave field whose phase is chosen so that its effective field in the rotating frame is aligned with the prepared transverse spin.
In the rotating frame, this drive defines a new quantization axis and splits the dressed spin states by the Rabi frequency $\Omega_R$.
If the spin is prepared parallel to this effective field, coherent Rabi precession is suppressed and the spin is instead ``locked'' along the drive axis.
Noise at frequencies near $\Omega_R$ can then induce transitions between the dressed states, causing the locked spin component to decay with a characteristic time $T_{1\rho}$ when fitted to an equation of the form \eqref{eq:PL_decay_curve}.
Thus $T_{1\rho}$ probes a spectral window between low-frequency dephasing measurements and high-frequency laboratory-frame $T_1$ relaxometry
\cite{
    rosskopf2014investigation,
    mzyk2022relaxometry}.
%

In all MW-controlled protocols, the amplitude must be chosen within a practical operating window.
The time for a resonant drive with Rabi frequency $\Omega_R$ to produce a rotation angle of $\theta$ is $t_\theta=
\theta/\Omega_R$.
Because the spin has a transverse component during these rotations, the pulses should be short compared with the dephasing time, setting a lower bound on $\Omega_R$ for high-fidelity rotations.
At the same time, using too large a MW amplitude can broaden the driven transition, reduce spectral selectivity, off-resonantly drive nearby hyperfine lines or other spin transitions, and introduce microwave heating or other technical artifacts.
A useful estimate for the onset of power broadening comes from the steady-state Bloch response of a driven two-level transition (i.e., the imaginary part of the AC susceptibility), given by
\begin{align}
\chi''\propto \frac{1}{1+\Delta^2T_2^2+\Omega_R^2T_1T_2}
\end{align}
and so the linewidth increases once the saturation parameter $\Omega_R^2T_1T_2$ becomes appreciable.
Thus, microwave power is typically chosen so that the pulses are fast enough to avoid appreciable dephasing, but not so strong that power broadening, off-resonant excitation, or drive-induced artifacts dominate
\cite{
    degen2017quantum,
    barry2020sensitivity,
    mzyk2022relaxometry}.

\subsection{NV Spin Hamiltonian and Transition Frequencies}

The role of the unperturbed spin-defect Hamiltonian in relaxometry is to determine the sensor eigenstates and transition frequencies, which set both the noise frequencies sampled by the defect and the microwave frequencies required for coherent control.
Excluding magnetic field noise and other effects, a minimal unperturbed Hamiltonian for the NV$^-$ ground state is
\begin{align}
\mathcal{H}/\hbar = D S_z^2 + \gamma_e \vb{B}\cdot\vb{S},
\end{align}
where $D$ is the zero-field splitting, $\gamma_e=2\pi\times28$\,GHz/T is the electron gyromagnetic ratio 
\cite{
    doherty2013nitrogen,
    rondin2014magnetometry,
    degen2017quantum}. 
The spin operator $\vb{S}=(S_x,S_y,S_z)$ denotes the spin-1 matrices acting on the triplet manifold and $\vb{B}$ is the static magnetic field.

When the applied field is approximately aligned with the NV axis, the eigenstates remain close to the $m_s=0,\pm1$ basis and the transition frequencies are approximately
\begin{align}\label{eq:ODMR_frequencies}
\omega_{0\leftrightarrow\pm1} = D \pm \gamma_e B_\parallel \,.
\end{align}
Here $B_\parallel=\vb{B}\cdot\hat{\vb{n}}_{\mathrm{NV}}$ is the projection of the static magnetic field along the NV symmetry axis $\hat{\vb{n}}_{\mathrm{NV}}$. 
The frequencies \eqref{eq:ODMR_frequencies} are known as the ODMR frequencies, due to the observed decrease in PL contrast when the MW drive frequency is swept through the resonance set by the applied magnetic field. 
Thus changing the static magnetic field tunes the frequency at which the NV samples transverse magnetic noise.
This frequency selectivity is the basis of both conventional $T_1$ relaxometry and field-dependent cross-relaxometry which will be discussed in more detail in Sec.~\ref{sec:T1_Relaxometry_Spectroscopy}.

Strain and off-axis fields $\vb{B}_\perp$ mix the $m_s=\pm1$ states and can produce avoided crossings, especially near level anticrossings where two spin levels become nearly degenerate, such as the ground-state level anticrossing (GSLAC) near $\gamma_e B_\parallel \approx D$.
Hyperfine couplings further split each electronic transition into multiple hyperfine lines, typically separated by a few MHz for the host nitrogen and for nearby strongly coupled $^{13}$C spins \cite{doherty2013nitrogen,degen2017quantum}.

In practice, relaxometry often treats the NV as an effective two-level system: either $\{\ket{0},\ket{-1}\}$ or $\{\ket{0},\ket{+1}\}$. 
This is justified when one transition is spectrally isolated and the other is far detuned; the neglected level then only contributes weakly via second-order processes \cite{degen2017quantum,barry2020sensitivity}. 
However, near level anticrossings all three levels and coupled nuclear spins can participate, and a full spin-1 (or spin-1 + nuclear spin) treatment is necessary to interpret $T_1$ and cross-relaxation features 
\cite{wood2017microwave,broadway2018quantum}.

\subsection{$T_1$ Relaxometry as Magnetic Noise Spectroscopy}\label{sec:T1_Relaxometry_Spectroscopy}

We now connect the measured relaxation rate to the magnetic noise produced by the environment.
The key idea is that $T_1$ relaxation is driven by magnetic-field fluctuations transverse to the sensor quantization axis.
These transverse fluctuations contain Fourier components at many frequencies, but only components near the spin transition frequency can efficiently drive transitions between spin states.
Thus, $T_1$ relaxometry converts magnetic noise at a particular frequency into a measurable population decay rate.

To make this connection explicit, consider an effective two-level system formed by $\ket{0}$ and $\ket{1}$, where $\ket{1}$ denotes one of the $m_s=\pm1$ states.
The two-level Hamiltonian can be written as
\begin{align}
\mathcal{H}_0
=
\frac{\hbar\omega_s}{2}\sigma_z,
\end{align}
where $\omega_s$ is the sensor transition frequency and $\sigma_z$ is the Pauli matrix.
Time-dependent magnetic noise enters as a perturbation,
\begin{align}
\mathcal{V}(t)/\hbar
=
\gamma_e \delta\vb{B}(t)\cdot\vb{S},
\end{align}
where $\delta\vb{B}(t)$ is the fluctuating magnetic field at the defect location.
The transverse components $\delta B_\perp(t)=[\delta B_x(t),\delta B_y(t)]$ couple the two spin states and can drive population relaxation.
The longitudinal component $\delta B_z(t)$ shifts the transition frequency and therefore primarily contributes to dephasing rather than to $T_1$ relaxation.

The magnetic noise PSD between Cartesian components $\alpha$ and $\beta$ is the Fourier transform of the autocorrelation function:
\begin{align}\label{eq:magnetic_noise_psd}
S_{B_\alpha B_\beta}(\omega)
=
\int_{-\infty}^{\infty}
\left\langle
\delta B_\alpha(0)\delta B_\beta(t)
\right\rangle 
e^{i\omega t}\dd{t},
\end{align}
where $t$ is the time delay and $\langle\,\cdot\,\rangle$ denotes an ensemble or time average over fluctuations.
For $T_1$ relaxometry, the relevant quantity is the transverse noise PSD evaluated at the sensor transition frequency, $S_{B_\perp}(\omega_s)$.
Here $S_{B_\perp}$ denotes the appropriate combination of transverse noise components, for example $S_{B_xB_x}+S_{B_yB_y}$ up to convention-dependent factors.
In the simple example of a magnetic field fluctuating according to a random telegraph process, the associated autocorrelation function is exponential
\begin{align}
\left\langle
\delta B_\perp(0)\delta B_\perp(t)
\right\rangle
=
\left\langle
\delta B_\perp^2
\right\rangle
e^{-|t|/\tau_c},
\end{align}
where $\tau_c$ is the correlation time. The corresponding PSD is therefore Lorentzian
\begin{align}
S_{B_\perp}(\omega)
=
\frac{
2\left\langle
\delta B_\perp^2
\right\rangle \tau_c
}{
1+\omega^2\tau_c^2
}
\end{align}
centered at zero frequency with linewidth set by $1/\tau_c$, shown schematically in Fig.~\ref{fig:theory}(b).
A longer correlation time produces a narrower spectrum, while shorter produces more broadband noise.
For a sensor with a finite $\omega_s$, the value of $\tau_c$ thus changes how much spectral density lies at $\omega_s$ and in turn changes the relaxation rate, which is relevant for samples dominated by low frequency noise.
If the fluctuating source has a characteristic oscillation or transition frequency $\omega_t$, the autocorrelation can instead contain an oscillatory factor, for example
\begin{align}
\left\langle
\delta B_\perp(0)\delta B_\perp(t)
\right\rangle
\propto
e^{-|t|/\tau_c}\cos(\omega_t t),
\end{align}
which produces Lorentzian peaks centered near $\omega=\pm\omega_t$.
In that case, the sensor relaxation rate is enhanced when the transition frequency $\omega_s$ overlaps one of these noise peaks, as occurs in cross-relaxometry with nearby spins or resonant magnonic modes.

Within Bloch--Redfield theory or Fermi's Golden Rule, the transition rates are proportional to the noise PSD at the transition frequency.
For the effective two-level system,
\begin{align}
\Gamma_{0\rightarrow1}
&\propto
\gamma_e^2 S_{B_\perp}(-\omega_s)
\\
\Gamma_{1\rightarrow0}
&\propto
\gamma_e^2 S_{B_\perp}(+\omega_s).
\end{align}
The two signs reflect that, in a quantum description, upward and downward transitions sample positive- and negative-frequency components of the bath spectrum.
For classical high-temperature noise, the PSD is effectively symmetric, $S_{B_\perp}(+\omega_s)\approx S_{B_\perp}(-\omega_s)$.
For low-temperature or genuinely quantum noise sources, this symmetry need not hold, and detailed-balance factors may become important
\cite{
degen2017quantum,
mzyk2022relaxometry}.

The measured longitudinal relaxation rate is the sum of the transition rates that redistribute population between the two levels:
\begin{align}\label{eq:gamma1_transition_rates}
\Gamma_1
=
\Gamma_{0\rightarrow1}
+
\Gamma_{1\rightarrow0}.
\end{align}
In the common classical-noise limit, this gives the central relaxometry relation
\begin{align}\label{eq:gamma1_psd_relation}
\Gamma_1^{\mathrm{sample}}
\propto \gamma_e^2
S_{B_\perp}^{\mathrm{sample}}(\omega_s),
\end{align}
where the exact prefactor is convention-dependent---containing spin matrix elements, geometric factors, and the precise definition of the PSD.
Equation \eqref{eq:gamma1_psd_relation} implies that the total $\Gamma_1=1/T_1$ is increased when the transverse magnetic noise PSD at the sensor transition frequency increases.
This is the central result of $T_1$ relaxometry: a population decay measurement gives a local, frequency-selective measurement of magnetic noise.

For the case of cross-relaxation between a sensor spin coupled to a target spin or mode, the resonant part of the interaction can be written as a flip-flop Hamiltonian,
\begin{align}\label{eq:flip-flop_Hamiltonian}
\mathcal{H}_{\mathrm{ff}}/\hbar
=
\frac{\Delta_\mathrm{cr}}{2}
\left(\sigma_z^s - \sigma_z^t\right) + J\left(\sigma_+^s\sigma_-^t + \sigma_-^s\sigma_+^t\right),
\end{align}
where $\Delta_\mathrm{cr}=\omega_s(B)-\omega_t(B)$ and $J$ is an effective coupling strength 
\cite{
    wood2016wide,
    wood2017microwave,
    broadway2018quantum}.
Indices $s$ and $t$ refer to sensor and target spins respectively, with $\sigma_\pm$ being ladder operators.
The resulting cross-relaxation contribution often takes the Lorentzian form 
\cite{
    wood2016wide,
    broadway2018quantum,
    mignon2023fast}
\begin{align}\label{eq:cross_relaxation_rate}
\Gamma_{1}^{\mathrm{cr}}(B)
\propto
\frac{
J^2\tau_c
}{
1+\Delta_\mathrm{cr}^2\tau_c^2
}.
\end{align}
Thus cross-relaxometry appears experimentally as a resonant enhancement of the sensor relaxation rate as a function of magnetic field.
The amplitude reflects the coupling strength, target density, and sensor--sample geometry, while the linewidth reflects the target correlation time together with any additional inhomogeneous broadening.

\subsection{Spectral Windows of Spin Relaxometry}

The $T_1$ measurement described above is one example of a broader principle: different control protocols make the defect sensitive to different parts of the environmental noise spectrum.
A useful way to organize these protocols is through the reduced density matrix of an effective two-level spin system,
\begin{align}\label{eq:density_matrix}
\rho
=
\begin{pmatrix}
\rho_{00} & \rho_{01} \\
\rho_{10} & \rho_{11}
\end{pmatrix}.
\end{align}
The diagonal elements describe spin populations, while the off-diagonal elements describe phase coherence between the two spin states.
Longitudinal relaxation changes the populations and is characterized by $T_1$, whereas dephasing suppresses the off-diagonal coherences and is characterized by $T_2^\ast$, $T_2$, or related coherence times depending on the control sequence.
Thus, in the broad sense used here, spin relaxometry includes both population-relaxation measurements and coherence-decay measurements.

%

In a simple Markovian description, the population difference and coherence obey Bloch-type equations of motion:
\begin{align}
\frac{d}{dt}
\left(
\rho_{11}-\rho_{00}
\right)
&=
-\frac{
\left(\rho_{11}-\rho_{00}\right)
-
\left(\rho_{11}^{\mathrm{eq}}-\rho_{00}^{\mathrm{eq}}\right)
}{T_1}, \label{eq:density_matrix_diag_EOM}
\\
\frac{d\rho_{01}}{dt}
&=
-\left(
i\omega_s
+
\frac{1}{T_2}
\right)
\rho_{01}\,. \label{eq:density_matrix_offdiag_EOM}
\end{align}
The homogeneous coherence time $T_2$ satisfies
\begin{align}\label{eq:T2_T1_Tphi}
\frac{1}{T_2}
=
\frac{1}{2T_1}
+
\frac{1}{T_\phi},
\end{align}
where $T_\phi$ is the pure-dephasing time.
This relation \eqref{eq:T2_T1_Tphi} shows that population relaxation contributes to coherence decay, but additional longitudinal noise can dephase the spin without changing its population.
For a two-level system in the high-temperature limit, thermal equilibrium is given by the fully mixed state $\rho^\mathrm{eq}=\mathbbm{1}_2/2$, where $\mathbbm{1}_2$ is the $2\times2$ identity matrix. 
Equations \eqref{eq:density_matrix_diag_EOM} and \eqref{eq:density_matrix_offdiag_EOM} then show that population relaxation drives the diagonal elements of $\rho$ toward equal occupation, while dephasing suppresses the off-diagonal coherences.
%

Different measurements probe different elements of this density matrix and different frequency components of the noise.
In a $T_1$ experiment, the sensor samples transverse magnetic noise near the transition frequency $\omega_s$ through changes in the diagonal populations.
In contrast, Ramsey, echo, and dynamical-decoupling measurements probe the decay of the off-diagonal coherence $\rho_{01}$ caused primarily by longitudinal noise.
Spin-locking probes relaxation in a driven rotating-frame basis, where the relevant splitting is set by the Rabi frequency $\Omega_R$.
These measurements can therefore be viewed as complementary spectral windows on the same fluctuating environment.

In measurements such as Ramsey, spin echo, and dynamical decoupling, the coherence is given by
\begin{align}
\rho_{01}(\tau) 
\propto
e^{-\chi(\tau)},
\end{align}
where $\chi(\tau)$ is the accumulated noise-induced phase variance after an evolution time $\tau$.
For dephasing caused by magnetic noise along the sensor quantization axis, one commonly writes
\begin{align}\label{eq:filter_function_dephasing}
\chi(\tau)
\propto
\gamma_e^2
\int_0^\infty
S_{B_\parallel}(\omega)
\left|F(\omega,\tau)\right|^2 \dd{\omega},
\end{align}
where $S_{B_\parallel}(\omega)$ is the PSD of longitudinal magnetic noise and $F(\omega,\tau)$ is a filter function determined by the pulse sequence
\cite{
degen2017quantum,
barry2020sensitivity}.
The prefactor and normalization of $F(\omega,\tau)$ depend on convention, but the physical meaning is simple: the measured coherence decay is determined by the overlap between the environmental noise spectrum and the frequency filter imposed by the control sequence.

The simplest coherence measurement is a Ramsey sequence.
After the first $\pi/2$ pulse, the spin evolves freely and accumulates phase from fluctuations in the local transition frequency.
Because no refocusing pulse is applied, Ramsey measurements are strongly sensitive to static, quasi-static, and low-frequency noise.
This includes slowly varying magnetic fields, temperature- or strain-induced frequency shifts, and spatial inhomogeneity in ensembles.
The resulting decay time $T_2^\ast$ is therefore an inhomogeneous dephasing time: it reflects both true temporal fluctuations and static differences between repetitions or between defects in an ensemble.

A simple expression for $T_2^\ast$ can be obtained in the quasi-static Ramsey limit.
If the transition frequency varies between experimental repetitions by an amount $\delta\omega$ drawn from a Gaussian distribution with standard deviation $\sigma_\omega$, then the normalized Ramsey coherence is
\begin{align}
W(\tau)
=
\left\langle e^{-i\delta\omega \tau}\right\rangle
=
\exp\left[
-\frac{1}{2}\sigma_\omega^2\tau^2
\right].
\end{align}
Writing this decay as $\exp[-(\tau/T_2^\ast)^2]$ gives
\begin{align}
T_2^\ast
=
\frac{\sqrt{2}}{\sigma_\omega}.
\end{align}
Thus $T_2^\ast$ is set by the width of the quasi-static distribution of transition frequencies, rather than by the noise PSD at a single finite frequency.
For magnetic dephasing noise, $\sigma_\omega=\gamma_e\sigma_{B_\parallel}$, where $\sigma_{B_\parallel}$ is the standard deviation of the longitudinal magnetic field projected along the sensor axis.

A Hahn-echo sequence adds a $\pi$ pulse halfway through the evolution.
This pulse reverses phase accumulation from static or slowly varying detuning, causing quasi-static frequency shifts to refocus at the end of the sequence.
As a result, echo measurements suppress the low-frequency noise that dominates Ramsey decay and instead become sensitive to noise that changes appreciably during the sequence.
More generally, multi-pulse dynamical-decoupling sequences use repeated $\pi$-pulses to create a filter function with passbands at frequencies set approximately by the pulse spacing.
By changing the interpulse spacing, one can shift these passbands and perform noise spectroscopy in the kHz--MHz range, depending on the available coherence time and pulse fidelity
\cite{
degen2017quantum,
barry2020sensitivity}.
Thus, the measured coherence time is not a single intrinsic material-dependent constant; it depends on both the spin environment as well as the pulse sequence used to define and measure it.

Spin-locking provides another spectral window.
In a $T_{1\rho}$ measurement, the spin is first rotated into the transverse plane and then held by a continuous resonant microwave drive.
In the rotating frame, the relevant energy splitting is set by the Rabi frequency $\Omega_R$, rather than by the laboratory-frame transition frequency $\omega_s$.
Noise near $\Omega_R$ can therefore drive relaxation between the dressed spin states, causing the spin-lock contrast to decay with a characteristic time $T_{1\rho}$.
Because $\Omega_R$ is typically much smaller than $\omega_s$, spin-lock relaxometry probes noise at intermediate frequencies, often between echo-based low-frequency dephasing measurements and laboratory-frame $T_1$ relaxometry
\cite{
mzyk2022relaxometry,
rosskopf2014investigation}.

A useful qualitative hierarchy is therefore:
\begin{align}
T_2^\ast &: \text{quasi-static/low-frequency dephasing noise}, \nonumber \\
T_2 \text{ / DD} &: \text{noise selected by pulse-defined filter functions}, \nonumber \\
T_{1\rho} &: \text{rotating-frame noise near } \Omega_R, \nonumber \\
T_1 &: \text{transverse noise near } \omega_s .  \nonumber
\end{align}

The practical takeaway is that these methods are not competitors but complementary probes of the same environment.
By combining Ramsey, echo, dynamical-decoupling, spin-locking, and $T_1$ measurements---possibly at multiple sensor depths, magnetic fields, and microwave powers---one can sample magnetic noise over many orders of magnitude in frequency.
The resulting spectral information can help separate different physical mechanisms, such as surface spins, phonons, magnons, electrical currents, nuclear spins, and superconducting fluctuations
\cite{
degen2017quantum,
mzyk2022relaxometry,
takei2024detecting}.
The next step is then to relate the local magnetic noise PSD sensed by the defect to the microscopic response functions of the sample.

\subsection{Relating Magnetic Noise to Material Response}

To extract material parameters from the relaxation rate, one must take one further step: connect the local magnetic noise $S_B(\omega)$ sensed by the defect to the microscopic degrees of freedom in the sample.
This is the central modeling problem in quantitative relaxometry.

This connection usually has two conceptually distinct pieces.
First, one needs a description of the intrinsic fluctuations in the material, such as spin, magnetization, current, quasiparticle, or vortex fluctuations.
These are encoded in correlation functions or, equivalently, in dynamical response functions such as magnetic susceptibilities or conductivities.
Second, one needs a geometrical propagator that describes how those fluctuating degrees of freedom generate magnetic fields at the defect position.
Thus the measured noise is not simply a property of the material alone, but of the combined material--geometry--sensor system.

The local magnetic noise PSD can be written as a source correlation function propagated to the sensor:
\begin{align}\label{eq:generic_source_to_field}
&S_{B_iB_j}(\vb{r}_s,\omega)
= \nonumber \\
&\iint
G_{i\alpha}(\vb{r}_s,\vb{r})\,
G_{j\beta}(\vb{r}_s,\vb{r}')\,
S_{X_\alpha X_\beta}(\vb{r},\vb{r}',\omega)\,\dd^3 r\,\dd^3 r',
\end{align}
where $\vb{r}_s$ is the sensor position, $\vb{r}$ and $\vb{r}'$ are positions within the fluctuating sample or source volume, and the integrals run over that source volume.
The Latin indices $i$ and $j$ label magnetic-field components at the sensor, while the Greek indices $\alpha$ and $\beta$ label components of the fluctuating source variable.
Repeated Greek indices are summed over Cartesian components.
Here $X_\alpha(\vb{r},t)$ denotes the relevant fluctuating source component at position $\vb{r}$, $S_{X_\alpha X_\beta}(\vb{r},\vb{r}',\omega)$ is the frequency-domain two-point correlation function of the source, and $G_{i\alpha}(\vb{r}_s,\vb{r})$ is the magnetostatic Green's function that converts a fluctuation of source component $\alpha$ at position $\vb{r}$ into magnetic-field component $i$ at the sensor.
The two Green's functions propagate correlated source fluctuations at $\vb{r}$ and $\vb{r}'$ to the sensor, producing the magnetic-field correlation between components $B_i$ and $B_j$.
For a magnetic sample, $X_\alpha$ may be the magnetization $M_\alpha$.
For a conductor it may be the current density $J_\alpha$, and for a superconductor it may include current, quasiparticle, or vortex degrees of freedom.
The measured $T_1$ rate then depends on the transverse projection of $S_{B_iB_j}(\vb{r}_s,\omega)$ evaluated at $\omega=\omega_s$.

Equation~\eqref{eq:generic_source_to_field} highlights an important point: the sensor does not measure all sample fluctuations equally.
The Green's function contains a strong spatial filter.
For near-field magnetic noise from a planar sample, Fourier components with in-plane wavevector $q=\abs{\vb{q}}$ are typically suppressed with increasing sensor height by factors that scale roughly as $e^{-qh}$ or $e^{-2qh}$ in the noise power.
As a result, fluctuations with wavelengths much shorter than the sensor--sample distance contribute weakly to the measured signal.
The NV depth, sample thickness, sensor orientation, and sample geometry therefore determine which parts of the material fluctuation spectrum are visible to the measurement.

For spin systems, such as paramagnets, ordered magnets, or magnetic insulators, the relevant fluctuating source is usually the local spin or magnetization density.
One often starts from the magnetization correlation function $S_{M_\alpha M_\beta}(\vb{q},\omega)$ or from the imaginary part of the dynamical magnetic susceptibility $\chi_{\alpha\beta}''(\vb{q},\omega)$.
At temperature $T$, the fluctuation--dissipation theorem relates these quantities schematically as
\begin{align}
S_{M_\alpha M_\beta}(\vb{q},\omega)
\propto
\frac{\chi_{\alpha\beta}''(\vb{q},\omega)}{1-e^{-\hbar\omega/k_BT}},
\end{align}
up to convention-dependent factors associated with the definition of the unsymmetrized or symmetrized spectrum.
The local magnetic noise PSD at the defect is then obtained by propagating these magnetization fluctuations through the appropriate magnetostatic kernel.
This framework is useful for describing relaxometry from paramagnetic spins, spin waves, magnons, and fluctuating magnetic domains
\cite{
degen2017quantum,
mccullian2020broadband,
takei2024detecting}.

For conducting samples, the relevant fluctuating sources are currents $\vb{J}(\vb{r},t)$.
These currents produce magnetic fields through the Biot--Savart law, and their fluctuations are connected to the conductivity tensor by the fluctuation--dissipation theorem.
Schematically,
\begin{align}
S_{J_\alpha J_\beta}(\vb{q},\omega)
\propto
\frac{\mathrm{Re}\left[
\sigma_{\alpha\beta}(\vb{q},\omega)\right]}{1-e^{-\hbar\omega/k_BT}}.
\end{align}
In the classical low-frequency limit, $\hbar\omega\ll k_BT$, this reduces to Johnson--Nyquist noise proportional to $k_BT\,\mathrm{Re}\left[\sigma_{\alpha\beta}\right]$ where $\sigma_{\alpha\beta}(\vb{q},\omega)$ is the nonlocal conductivity tensor.
The resulting magnetic noise at the NV is then found by propagating the fluctuating current density to the sensor position.
Analytic expressions exist for common geometries, such as conducting half-spaces and thin films, and show how $\Gamma_1$ depends on temperature, conductivity, film thickness, skin depth, and sensor--sample distance
\cite{
kolkowitz2015probing,
ariyaratne2018nanoscale,
degen2017quantum}.

For superconductors, the same conceptual structure applies, but the relevant electrodynamics can be more complex.
Meissner screening, quasiparticles, collective modes, and vortex motion can all modify the current fluctuations that produce magnetic noise at the sensor.
Depending on temperature, magnetic field, and sample geometry, superconductivity can suppress magnetic noise by screening currents, or enhance it through dissipative vortex or quasiparticle dynamics.
Thus, superconducting relaxometry requires a model for both the fluctuating degrees of freedom and the electromagnetic response that propagates those fluctuations to the defect
\cite{
liu2025quantum,
kelly2024superconductivity}.

In all of these cases, geometry and distance matter as much as the intrinsic material response.
Two samples with the same local susceptibility or conductivity can produce different magnetic noise PSDs at the sensor if their thickness, shape, orientation, or distance from the defect differs.
Conversely, changing the defect--sample separation can be used as a diagnostic tool, because different microscopic noise mechanisms often have different distance dependences.
Careful modeling of the source-to-field propagator, or at least robust scaling tests versus sensor--sample distance and sample geometry, is therefore essential for quantitative relaxometry
\cite{
degen2017quantum,
mzyk2022relaxometry,
kolkowitz2015probing,
mccullian2020broadband}.

The broad lesson is that relaxometry is local magnetic noise spectroscopy, not a direct measurement of a material response function by itself.
A measured change in $\Gamma_1$ must be interpreted through both the material fluctuation spectrum and the magnetic-field propagator connecting the sample to the defect.


\section{Experimental Applications of Spin Relaxometry}


\subsection{Condensed matter systems}

\subsubsection{Conductors}
In regular conductors, the magnetic noise in the system is dominated by fluctuations arising from thermally excited current, commonly described as Johnson-Nyquist noise.
The stochastic thermal currents produce short-range broadband magnetic fields that encompass the transition frequencies between the $m_s=0$ and $m_s=\pm 1$ states, thereby driving spin relaxation in nearby defects \cite{kolkowitz2015probing, ariyaratne2018nanoscale}.

\begin{figure*}[t]
    \centering
    \includegraphics[width=0.84\linewidth]{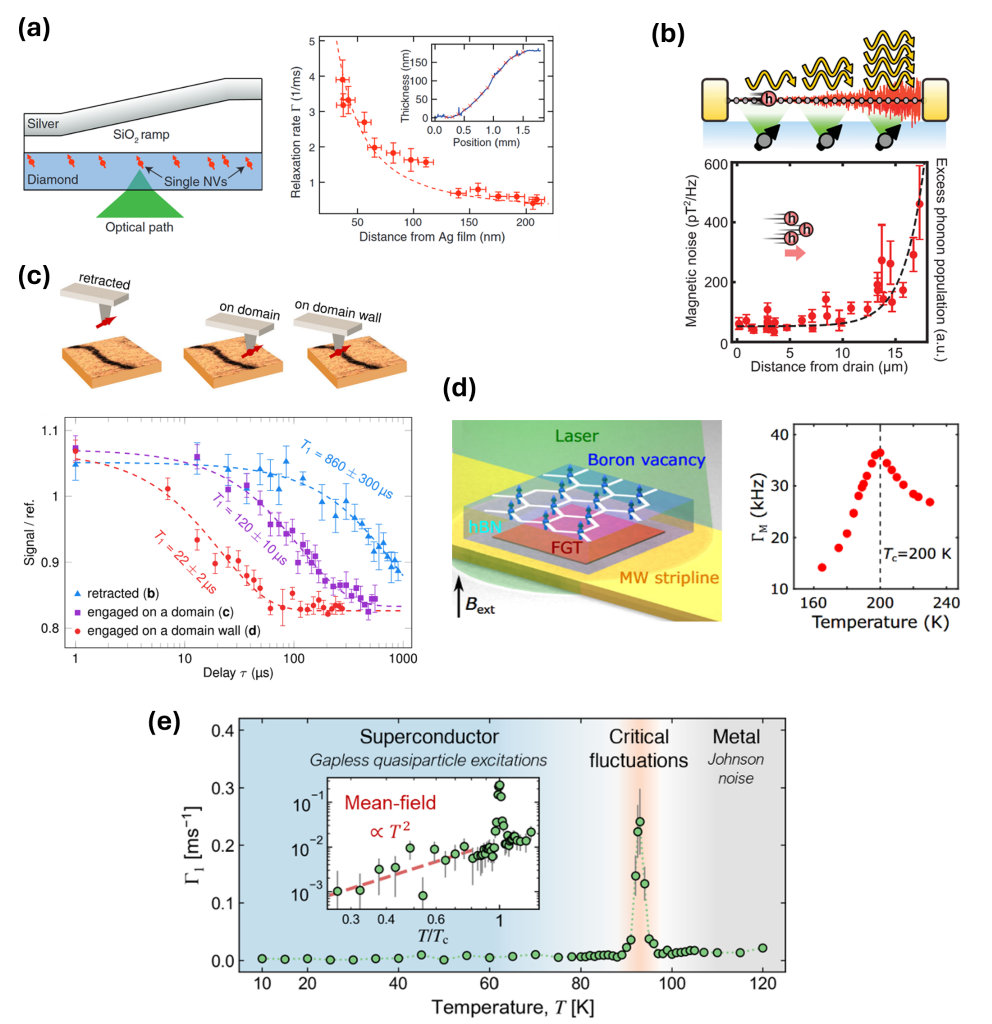}
    \caption{{\bf Experimental applications in condensed matter systems.}
    (a) Probing Johnson noise in metal using single NV centers.
    A layer of SiO$_2$ is grown on the diamond surface with gradually increasing thickness, followed by a 60\,nm silver film.
    The NV relaxation rate $\Gamma_1$ is measured as a function of distance from the silver film, scaling inversely with distance.
    Adapted from Ref.~\cite{kolkowitz2015probing}.
    (b) Investigating non-equilibrium dynamics in graphene.
    A hBN-encapsulated graphene device on diamond substrate (upper panel).
    The local magnetic noise, probed by the NV relaxation rate, is measured as a function of distance from the drain, consistent with the exponential growth of phonons (lower panel).
    Adapted from Ref.~\cite{andersen2019electron}.
    (c) Imaging domain walls in a synthetic antiferromagnet.
    The relaxation time $T_1$ of the NV center decreases dramatically when engaged on a domain wall, compared to on domain or retracted.
    Adapted from Ref.~\cite{finco2021imaging}.
    (d) Imaging the spin fluctuations in a van der Waals ferromagnet with boron vacancy centers in hBN.
    A Fe$_3$GeTe$_2$(FGT)/hBN van der Waals heterostructure is transferred on a gold microwave stripline (left panel).
    The temperature-dependent relaxation rate exhibits a peak amplitude at the phase transition temperature $T_c = 200$~K (right panel). 
    Adapted from Ref.~\cite{huang2022wide}.
    (e) Probing superconducting dynamics in a thin film Bi$_2$Sr$_2$CaCu$_2$O$_{8+\delta}$ (BSCCO).
    The relaxation rate $\Gamma_1$ of NV on BSCCO is measured as a function of temperature.
    In the absence of magnetic field, it reveals three distinct regimes: Johnson noise in metallic phase, nodal quasiparticles excitations deep in superconducting phase, and critical fluctuations near the phase transition.
    Adapted from Ref.~\cite{liu2025quantum}.
    }
    \label{fig:condensed}
\end{figure*}

In 2015, Kolkowitz \emph{et al.} employed shallow NV spins to detect field fluctuations originating from proximal silver films \cite{kolkowitz2015probing}.
By depositing a layer of SiO$_2$ between the diamond surface and silver film with a gradually increasing thickness, they measured the variations in the longitudinal spin relaxation rate $\Gamma_1$ with the NV-sample separation distance $d$ (Fig.~\ref{fig:condensed}a).
Combined with temperature control, they reached a model where the relaxation rate scales with the conductivity $\sigma$ of the metal and the inverse of the distance $d$, as $\Gamma_1 \propto T\sigma/d$.
This work established NV relaxometry as a non-invasive probe capable of measuring local electron transport without the need for electrical contacts.

Subsequent studies extended this technique to non-equilibrium transport dynamics in two-dimensional materials.
Andersen \emph{et al.}~(2019) utilized shallow NV centers to probe the local magnetic noise generated by high-mobility graphene devices driven into the non-linear transport regime \cite{andersen2019electron}.
By mapping the NV spin relaxation rate along the conduction channel, the authors revealed a spatial asymmetry in the current fluctuations that was inaccessible via global transport measurements.
The local noise was found to grow exponentially along the direction of carrier drift, with the spatial profile inverting upon reversal of the current or charge carrier sign (Fig.~\ref{fig:condensed}b).
These results demonstrate the unique capability of solid-state defects to image the buildup of collective excitations in mesoscopic systems.

\subsubsection{Ferromagnets and Antiferromagnets}
The application of NV relaxometry to magnetically ordered systems has diverged into two main streams: imaging spin textures that do not generated net stray field, and probing the critical dynamics near magnetic phase transitions.

\emph{Imaging Antiferromagnetic Textures} --- 
The spin textures such as domain walls, spin spirals, and skyrmions in antiferromagnets are notoriously difficult to image using conventional methods due to their vanishing net magnetization.
While these textures may not generate a static magnetic field, they are sites of localized spin dynamics, which couples to NV relaxation.
In 2021, Finco \emph{et al.} demonstrated an all-optical imaging mode using single NV center where the relaxation rate is mapped by recording the photoluminescence (PL) intensity under continuous laser illumination~\cite{finco2021imaging}.
At domain walls, a significantly enhanced spin relaxation is observed.
In contrast, at uniform antiferromagnetic domain region, less spin noise was detected due to the spin wave gap (Fig.~\ref{fig:condensed}c).
This work established relaxometry as a versatile tool for mapping materials where static stray fields are negligible.

\emph{Probing Phase Transitions} --- 
Near a continuous magnetic phase transition, such as the Curie temperature ($T_c$) of a ferromagnet or the Néel temperature ($T_n$) of an antiferromagnet, the correlation length of spin fluctuations diverges, manifesting as a dramatic increase in magnetic noise~\cite{li2025criticalFGT}.

Recent studies have utilized relaxometry to map the phase transition of various magnetic materials with nanoscale precision, especially in 2D van der Waals magnets~\cite{huang2022wide,ziffer2024quantum, wu2025nanoscale}.
For instance, Huang \emph{et al.} (2022) demonstrated nanoscale imaging of low-dimensional ferromagnetism in Fe$_3$GeTe$_2$/hBN van der Waals heterostructures~\cite{huang2022wide}.
By performing wide-field spin relaxometry on boron-vacancy centers in hBN, they observed a distinct peak in the relaxation rate around the Curie temperature (Fig.~\ref{fig:condensed}d).
This enhancement was attributed to the critical divergence of the longitudinal magnetic susceptibility near the phase transition.
These experiments highlight the capability of solid-state spins to act as local probes of criticality.


\subsubsection{Superconductors}
Superconductors represent a macroscopic quantum state of matter defined by zero electrical resistance and the perfect expulsion of magnetic fields.
%
%
The pronounced, highly localized magnetic responses inherent to superconductors make solid-state spin defects, particularly nitrogen-vacancy (NV) centers in diamond, an exceptionally promising platform for non-invasive quantum sensing.
Pioneering in 2011, Bouchard \emph{et al.} first deployed the NV magnetometry in close proximity to a superconducting surface~\cite{bouchard2011detection}.
By monitoring the frequency shifts of the NV magnetic resonance spectra, they observed the Meissner effect as the system was cooled below the transition temperature.
Later works by Thiel \emph{et al.} (2016)~\cite{thiel2016quantitative} and Schlussel \emph{et al.} (2018)~\cite{schlussel2018wide} have extended this effects, and achieved spatial imaging of individual superconducting vortices deep within the superconducting phase.
Whereas these studies primarily utilized NV centers to characterize the static magnetic features, recent advances have moved beyond static imaging to probing the dynamics of the superconducting condensate and its excitations via relaxometry.

The first experimental demonstration of this technique was achieved in a high-$T_c$ superconductor Bi$_{2}$Sr$_{2}$CaCu$_{2}$O$_{8+\delta}$ (BSCCO)~\cite{liu2025quantum}.
By transferring an exfoliated BSCCO flake onto the diamond surface hosting a shallow layer of NV centers, Liu \emph{et al.} (2025) utilized relaxometry to map distinct dynamical regimes across different phases.
At low temperatures ($T \ll T_c$), the authors identified a $\Gamma_1 \propto T^2$ power-law dependence, providing direct evidence for nodal quasiparticle excitations characteristic of $d$-wave pairing.
Approaching the critical temperature $T_c$, the relaxation rate exhibited a sharp divergence attributed to critical fluctuations of the superconducting order parameter (Fig.~\ref{fig:condensed}e).
Furthermore, under applied magnetic fields, the additional noise induced by diffusive motion of the vortex liquid exhibit a clear linear dependence on the applied field strength, enabling the quantitative extraction of vortex diffusivity.
These results establish quantum relaxometry as a versatile platform for resolving complex dynamical phenomena in correlated materials.

\subsection{Biological Applications}

A central goal in bio-sensing is quantitative, localized readout of physiologically relevant variables (ions, radicals, biomolecules, pH, temperature, etc.) inside living systems.
Conventional modalities often face tradeoffs among sensitivity, selectivity, invasiveness, and resolution, and many cannot access microscopical processes in single living cells under physiological conditions \cite{schirhagl2014nitrogen, aslam2023quantum}.

Solid state defect, with NV centers in particular, are attractive because they combine optical addressability at room temperature with excellent bio-compatibility from their chemical inertness.
The most popular platform for studying biological processes is nanodiamond/fluorescent nanodiamond (FND) for straightforward integration with biological systems.
FNDs can be easily dispersed in liquids, surface-functionalized, and internalized by cells.
At the same time, it can be also implemented with other modalities including scanning NV \cite{wang2019nanoscale}, wide-field NV layers, and etc. with varying spatial resolution, photon collection, and sample compatibility.

For a broader perspective on diamond NV–based biosensing beyond relaxometry, the reader is referred to recent of quantum sensors for biomedical applications and focused reviews on fluorescent nanodiamond/NV biosensing platforms \cite{aslam2023quantum, zhang2021toward, wu2022recent}. 
Here we narrow the discussion to relaxometry-driven biological readouts.

Many biological important signals are carried by unpaired electron spins, including radicals, transition-metal ions, and spin labels.
In these settings, $T_1$ relaxometry is particularly useful because fluctuating electron spins generate transverse magnetic noise that shortens $T_1$.
Relaxometry in cells focuses mapping the spin-fluctuation-driven changes in $T_1$ to infer concentrations of dynamics of paramagnetic species, including free radicals and paramagnetic ions.

\begin{figure*}[t]
    \centering
    \includegraphics[width=0.7\linewidth]{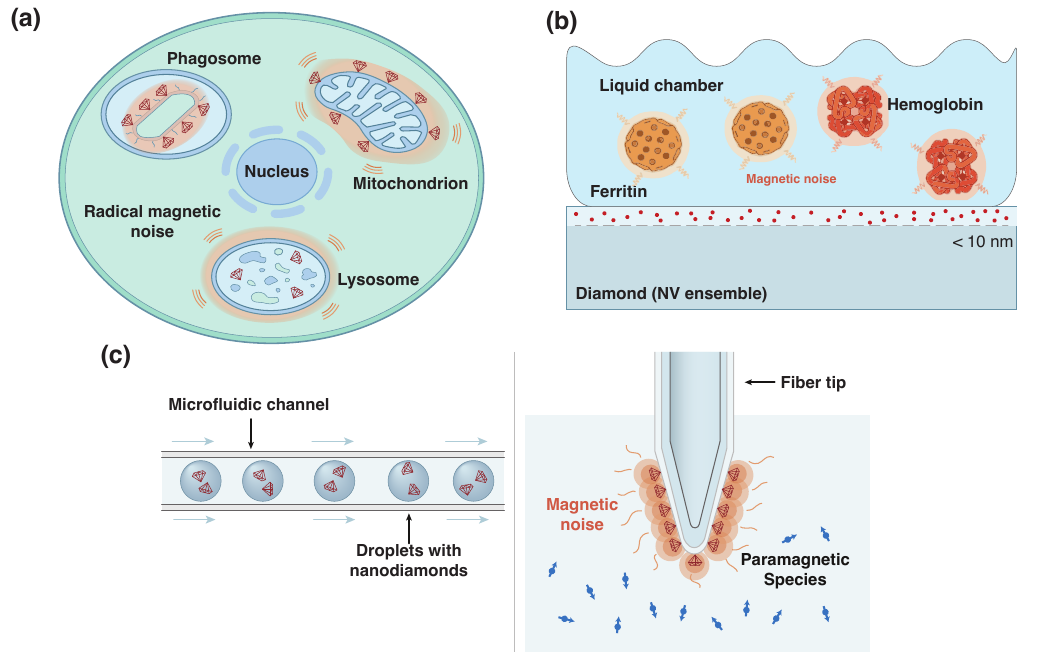}
    \caption{\textbf{Experimental applications in biological systems.}
    (a) Intracellular radical dynamics. Targeted fluorescent nanodiamonds enable nanoscale $T_1$ relaxometry inside living cells, allowing spatially and temporally resolved readout of radical-driven spin noise at specific intracellular locations, including bacterial surfaces, mitochondria, and lysosomes.
    (b) Chip-scale biomolecular assays. Planar diamond substrates with shallow near-surface NV ensembles provide quantitative relaxometry of paramagnetic biomolecules in controlled assay environments.
    (c) Integrated and scalable platforms. Microfluidic and fiber-integrated relaxometry platforms address throughput and accessibility by enabling signal modulation, background suppression, and remote sensing in complex biological fluids.
    }
    \label{fig:bio}
\end{figure*}

Since the 2022 review by Schirhagl \emph{et al.}~\cite{mzyk2022relaxometry}, the field has matured from detecting static concentrations of paramagnetic ions in buffer to resolving spatiotemporal dynamics in living systems.
The recent literature is characterized by a shift toward mapping non-equilibrium ``radical dialogues" in host-pathogen interactions, quantifying magnetic phase transitions in metalloproteins, and validating alternative host materials like silicon carbide (SiC) for physiological compatibility. Figure \ref{fig:bio} displays a summary of NV spin relaxometry used in bioscience.
In surveying these applications, it is useful to distinguish quantitatively calibrated assays from demonstrations of relative contrast under favorable conditions: in-vitro, chip-scale measurements currently provide the firmest numbers---ferritin detected down to ${\sim}7.5\,\mu$g/mL~\cite{grant2022nonmonotonic} and methemoglobin read out in sub-100\,pL volumes~\cite{lamichhane2024magnetic}---whereas several intracellular studies report relative $T_1$ changes with sensors conjugated directly to the target rather than absolute concentrations~\cite{wu2023diamond,wu2024nanoscale,fan2024quantum}. 
We carry this distinction through the subsections below.

\subsubsection{Intracellular Radical Dynamics and Metabolic Profiling}

The primary biological application of nanodiamond (ND) relaxometry remains the detection of reactive oxygen species (ROS), but recent work has moved beyond binary ``stress detection'' to resolve complex signaling timelines (Fig.~\ref{fig:bio}a).

\emph{Mapping the ``Radical Dialogue'' in Infection.} A persistent challenge in immunology is distinguishing between the host's oxidative attack and the pathogen's antioxidative defense.
In 2023, Wu \emph{et al.}~utilized fluorescent nanodiamonds (FNDs) conjugated directly to the surface of Staphylococcus aureus to probe this interaction inside macrophages \cite{wu2023diamond}.
Relaxometry measurements revealed a biphasic response: an initial $T_1$ reduction corresponding to the macrophage oxidative burst, followed by a recovery of $T_1$ specifically at the bacterial surface starting 80 minutes post-infection.
Quantitatively, $T_1$ recovered to ${\sim}170\,\mu$s at the bacterial surface; because the FNDs were conjugated directly to the surface, the measurement reports relative radical dynamics rather than an absolute radical concentration or detection limit.
This provided the first direct, nanoscale evidence of the pathogen actively scavenging radicals to evade immune clearance, a localized survival mechanism invisible to bulk ROS assays.

\emph{Organelle-Specific Metabolic Footprinting.} 
Targeted sensing has revealed that delivery vectors themselves alter metabolic states, a confound often overlooked in standard toxicity assays. 
Wu \emph{et al.}~(2024) demonstrated that the mitochondrial targeting ligand Triphenylphosphonium (TPP) inherently downregulates intracellular radical production, whereas the signaling peptide Somatostatin (SST) upregulates it \cite{wu2024nanoscale}.
Similarly, Fan \emph{et al.}~(2024) applied antibody-targeted FNDs to track real-time mitochondrial radical bursts in keratinocytes during UVB exposure \cite{fan2024quantum}.
These studies highlight the necessity of controlling for the ``oxidative footprint" of the sensor's own targeting moiety to avoid misinterpretation of metabolic data.

\emph{Mechanistic Chemistry In Situ.} Relaxometry has also been applied to elucidate intracellular reaction mechanisms. 
Lu \emph{et al.}~(2024) coated NDs with a thin layer of eumelanin—a pigment with complex radical chemistry—and internalized them into lysosomes \cite{lu2024unraveling}. 
The $T_1$ response quantified the formation of semiquinone radicals via a comproportionation reaction, which was found to be strictly pH-dependent. 
By correlating the radical count with theoretical models, the sensors acted as transducers, mapping the lysosomal pH via the radical proxy.

\subsubsection{Chip-Scale Interfaces and Biomolecular Assays}
Planar diamond substrates with shallow NV ensembles (<\,10\,nm depth) have matured into quantitative platforms for analyzing metalloproteins (Fig.~\ref{fig:bio}b), though recent physical insights demand updated calibration protocols. 
Freire-Moschovitis \emph{et al.}~(2023) overturned the assumption that physiological salts primarily add magnetic noise; they found that diamagnetic electrolytes actually increase $T_1$ by stabilizing surface charge states via electric double layer formation \cite{freire2023role}. 
This ``diamagnetic effect'' requires that bio-assays rigorously control ionic strength to avoid artifacts. 
In applications, Grant \emph{et al.}~(2023) utilized relaxometry for ``magnetic phasemetry'' of ferritin, revealing a nonmonotonic noise scaling that signifies a structural transition from disordered iron clusters to a crystalline core \cite{grant2022nonmonotonic}. 
Clinically, Lamichhane \emph{et al.}~(2024) demonstrated the quantification of methemoglobin in <\,100\,pL volumes, establishing a linear response curve suitable for rapid, micro-scale blood analysis \cite{lamichhane2024magnetic}.
 
\subsubsection{Integrated Platforms: Microfluidics and Fibers}

To address the low throughput of single-particle tracking, engineering efforts have focused on integrating quantum sensors with scalable fluidics (Fig.~\ref{fig:bio}c). 
Sarkar \emph{et al.}~(2024) introduced a ``double lock-in'' scheme by encapsulating nanodiamonds in flowing picoliter droplets \cite{sarkar2024high}.
The periodic flow modulates the fluorescence signal, suppressing low-frequency background noise to levels comparable with bulk crystals and enabling high-throughput single-cell screening.
For remote sensing, Cheng \emph{et al.}~(2024) developed an all-fiber probe with nanodiamonds chemically anchored to a tapered tip \cite{cheng2024all}. 
This ``dip-stick'' geometry successfully detected pH changes and specific biomarkers in bulk fluids where optical access is otherwise restricted.

\subsubsection{Emerging Materials and In Vivo Translation}

While diamond remains the gold standard, we have seen significant milestones in alternative host materials and whole-organism imaging.

Bridging the gap to multicellular organisms, Fan \emph{et al.}~(2025) reported the first in vivo relaxometry mapping of oxidative stress in a Caenorhabditis elegans model of Huntington’s disease \cite{fan2025vivo}.
By microinjecting FNDs into specific tissues, they detected a statistically significant elevation of free radicals in body wall muscles expressing PolyQ aggregates compared to the intestine, directly linking protein aggregation to localized oxidative stress in a living animal.

Silicon carbide has emerged as a novel platform for biological applications due to its biocompatibility and industrial maturity. 
Li \emph{et al.}~(2025) demonstrated stable, room-temperature divacancy qubits in alkene-terminated SiC that function as bioinert sensors \cite{li2025non}.
Crucially, SiC defects emit in the near-infrared (NIR), offering superior tissue penetration compared to the visible emission of NV centers. 
Parallelly, hexagonal Boron Nitride (hBN) is being explored for its 2D nature, which allows defects to be atomically close to the target. Robertson \emph{et al.}~(2023) utilized hBN powder to detect paramagnetic Gd$^{3+}$ ions, proving that van der Waals materials can bring the sensor-target separation to the angstrom scale, maximizing dipolar coupling \cite{robertson2023detection}.

\subsubsection{Outlook and Bottlenecks}
Across reviews, the main bottlenecks are consistent: improving sensitivity and temporal resolution for dynamic measurements in biological environments, enabling programmable targeting to specific intracellular locations, and mitigating cross-talk among simultaneously varying parameters.

\subsection{Spin Dynamics: Cross-Relaxation, Diamond Surface, and Nuclear Spins}

\subsubsection{Cross-relaxation spectroscopy}
Cross-relaxation (CR) spectroscopy can be understood as a field-tunable, frequency-selective $T_1$ measurement: the NV longitudinal relaxation rate $\Gamma_1 \equiv 1/T_1$ samples the transverse magnetic noise spectral density near the NV transition frequency $\omega_{\mathrm{NV}}$ (see Sec.~\ref{sec:T1_Relaxometry_Spectroscopy} for theory). Sweeping the bias field $B_0$ tunes $\omega_{\mathrm{NV}}(B_0)$ and converts spectral features in the environment into a characteristic field dependence $\Gamma_1(B_0)$. When $\omega_{\mathrm{NV}}$ approaches a discrete bath resonance, dipolar-mediated exchange is enhanced and produces sharp signatures in $\Gamma_1$ (dips in $T_1$) whose linewidths are set by the effective broadenings of the NV and bath transitions \cite{hall2016detection}.

A canonical benchmark is the NV--P1 ``standard candle.'' Hall \emph{et al.}~demonstrated $T_1$-based electron spin resonance (ESR) spectroscopy by tuning an NV ensemble through the NV--P1 resonance near $B_0 \approx 512$\,G and observing a pronounced reduction of $T_1$, then using the field dependence to reconstruct the hyperfine-resolved ESR spectrum of the substitutional-nitrogen (P1) bath \cite{hall2016detection}.
Beyond its pedagogical value, this result established a practical workflow that recurs across modern CR experiments: (i) identify resonance fields where $\Gamma_1$ is enhanced, (ii) fit the dip features to obtain resonance positions and linewidths, and (iii) map those features back to bath transition frequencies and couplings.

CR features are also a useful handle in dense, interacting spin environments where multiple sub-ensembles coexist.
For example, P1-resolved relaxation signatures have been leveraged to separate frequency-selected channels within strongly interacting dipolar baths, allowing NV depolarization dynamics to report on which bath subgroups participate in resonant exchange \cite{zu2021emergent}.
Systematic ensemble measurements further clarified how these channels compete with phonons: Jarmola \emph{et al.}~mapped $T_1$ versus field and temperature and showed that sample-dependent cross-relaxation processes (including NV--NV and NV--P1 channels) can dominate low-temperature relaxation when phonon-assisted pathways are suppressed \cite{jarmola2012temperature}.
Complementary concentration-dependent studies established that increasing defect densities strengthens and broadens CR resonances, making $\Gamma_1(B_0)$ line shapes a quantitative diagnostic of paramagnetic-impurity density and disorder in NV ensembles \cite{mrozek2015longitudinal}.

Two practical features make CR relaxometry attractive for applications.
First, it is naturally sensitive to fast-fluctuating electron-spin targets whose noise power resides in the GHz band, and it can be deployed as a wide-band spectroscopy tool by tuning $\omega_{\mathrm{NV}}$ over a large frequency range via $B_0$ \cite{wood2016wide,hall2016detection}.
Second, spectral selectivity continues to improve as experiments and models better disentangle overlapping impurity signatures and microwave-driven artifacts in optically detected CR observables \cite{lazda2021cross}.

An important extension of CR physics appears near the NV GSLAC at $B_0 \approx 1024$\,G, where electron-nuclear mixing reshapes the relevant splittings and enables purely optical access to low-frequency magnetic noise.
Broadway \emph{et al.}~showed that, in this regime, microwave-free protocols can detect magnetic noise components from the MHz scale down toward the sub-MHz range, opening an all-optical route to CR-like sensitivity in a band that overlaps nuclear Larmor frequencies at modest fields \cite{broadway2016anticrossing}.

Finally, CR relaxometry is increasingly used as a \emph{readout layer} for optically ``dark'' spins in heterogeneous materials. Recent scanning-NV work demonstrated ESR detection of boron-vacancy defects ($V_{\mathrm{B}}^{-}$) in hBN by monitoring changes in the NV $T_1$ while tuning the NV transition through the NV--$V_{\mathrm{B}}^{-}$ cross-relaxation condition, thereby avoiding any requirement for direct fluorescence-based readout of the target defect \cite{melendez2025probing}. This hybrid 2D/3D strategy highlights a broader theme: once the NV is established as a calibrated relaxometric transducer, CR spectroscopy can be extended to material platforms where conventional optical or inductive ESR is impractical. Moreover, the scanning geometry enables sub-diffraction-limited, spatially resolved mapping of defect populations, providing a quantitative route to imaging defect density variations across heterogeneous 2D landscapes (Fig.~\ref{fig:spin}a).

\subsubsection{Surface spin noise and charge dynamics in shallow-NV relaxometry}

Shallow NV centers (typically $\lesssim$\,10\,nm below the diamond interface) are essential for nanoscale relaxometry because the signal from external targets rises rapidly as the NV is brought closer to the sample. In practice, the same proximity exposes the NV to a second, unavoidable ``target'': magnetic and electric-field fluctuations associated with the diamond surface. These surface-induced fluctuations can dominate both $T_1$ and $T_2$, setting a sensitivity floor that must be quantified and engineered around.

Multiple experiments show that relaxation and dephasing accelerate sharply as the NV approaches the surface, consistent with a surface-localized noise source rather than a bulk mechanism. Rosskopf \emph{et al.}~measured $T_1$, $T_{1\rho}$, and $T_2$ for very shallow NV centers (depth $\sim 5$\,nm) and observed relaxation-time reductions of up to $\sim$\,30 times relative to bulk, consistent with ubiquitous surface-associated magnetic impurities~\cite{rosskopf2014investigation}. Their analysis supports a dilute but consequential surface spin density (reported as $\sim$\,0.01 to $0.1\,\mu_B/\mathrm{nm}^2$) and fast surface-spin dynamics with a characteristic correlation time $\tau_c \approx 0.28$\,ns. 
In a complementary approach, Myers \emph{et al.}~used depth-calibrated NVs and dynamical decoupling to isolate surface-driven dephasing, with results consistent with a surface bath of electronic spins and a correlation rate on the order of $200$~kHz in the spectral window relevant to dephasing \cite{myers2014probing}. These two studies are consistent in the key applied message: the dominant noise is localized near the interface, and it is dynamical rather than purely static.

A practical point for relaxometry is that ``the correlation time'' extracted from an NV experiment depends on the frequency band being interrogated. 
Rosskopf \emph{et al.}~reported sub-nanosecond dynamics compatible with a broadband magnetic spectrum that can efficiently relax the NV at its transition frequency \cite{rosskopf2014investigation}. 
Romach \emph{et al.}~further emphasized that shallow NVs experience a structured spectrum: their analysis supports a double-Lorentzian form with a low-frequency component consistent with a surface electronic spin bath and an additional faster component attributed to surface-modified phononic coupling \cite{romach2015spectroscopy}. 
For applied work, this implies that improving $T_2$ does not automatically guarantee improved $T_1$, and vice versa. Both must be measured in the operating regime of the intended sensing protocol.

\begin{figure}[t]
    \centering
    \includegraphics[width=\columnwidth]{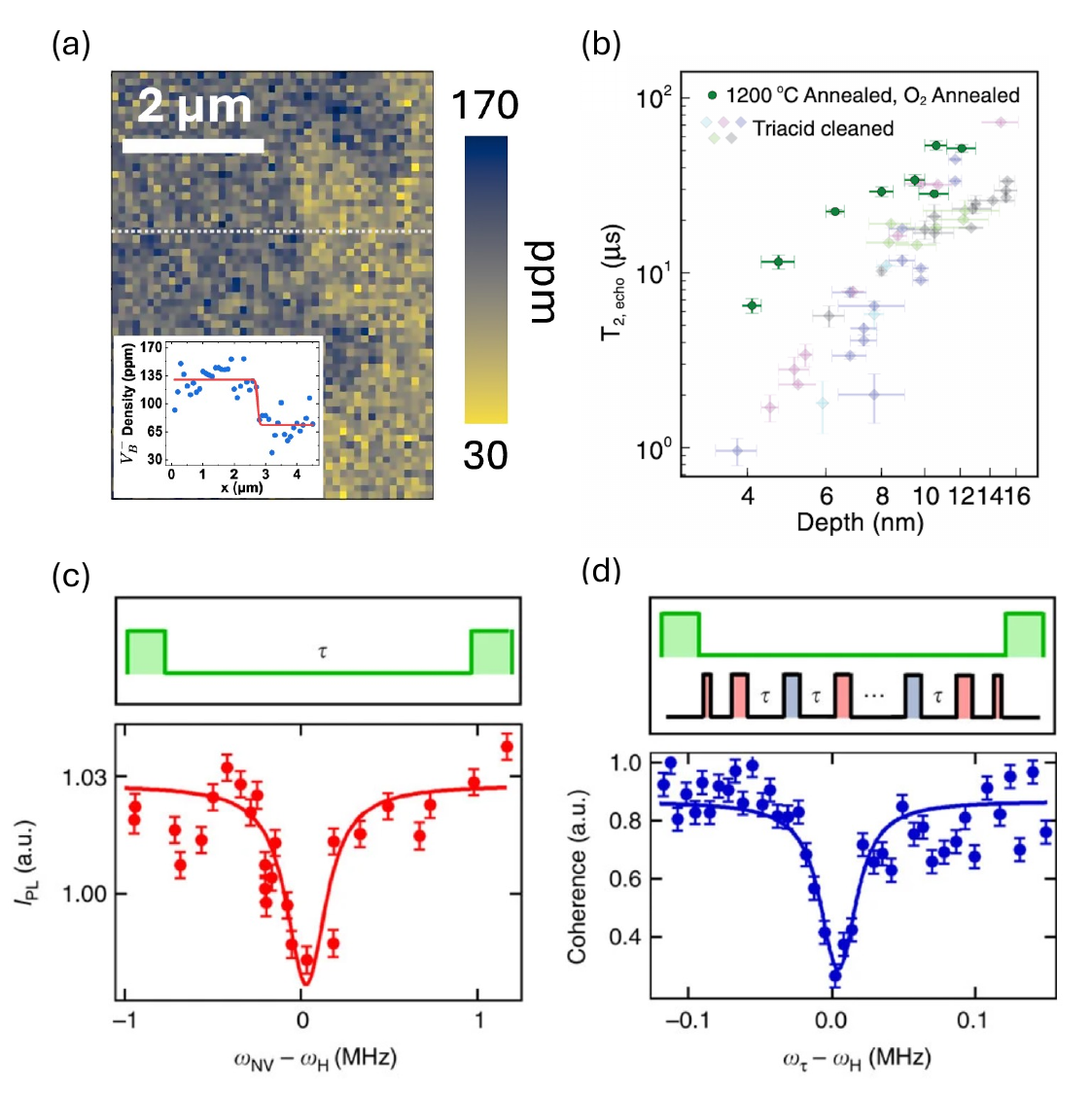}
    \caption{
    \textbf{Experimental applications in spin dynamics.} 
    (a) Defect density map of boron vacancies obtained using scanning NV cross-relaxometry. Inset: The profile of \vbm density as a function of position, over the line indicated by the white dashed line. Adapted from Ref.~\cite{melendez2025probing}.
    (b) Comparison of Hahn echo coherence times ($T_{2,\text{echo}}$) as a function of NV depth. The high-temperature and oxygen-annealed sample (colored markers) exhibits significantly improved coherence times at the same depths compared to those under a standard triacid-cleaned surface (grey markers). Adapted from Ref.~\cite{sangtawesin2019origins}.
    (c)-(d) Representative spectra from poly(methyl methacrylate) (PMMA) measured with a single $^{14}NV$ center using (c) microwave-free $T_1$ relaxometry and (d) an XY8-N dynamical-decoupling sequence (N=256 microwave $\pi-$pulses). The corresponding pulse sequences are shown schematically, with laser pulses in green and microwave pulses in red ($0^\circ$ phase) or blue ($90^\circ$ phase). Adapted from Ref.~\cite{wood2017microwave}.
    }
    \label{fig:spin}
\end{figure}

Attempts to reduce surface noise have shown that chemical termination alone is not the full story; surface morphology controls whether a termination is reproducible and stable. Sangtawesin \emph{et al.}~combined surface spectroscopy with single-NV measurements and demonstrated that a highly ordered, oxygen-terminated surface can suppress noise (Fig.~\ref{fig:spin}b), with shallow NV centers (within $10$\,nm) exhibiting coherence times extended by about an order of magnitude~\cite{sangtawesin2019origins}. 
In parallel, Stacey \emph{et al.}~provided evidence that diamond surfaces can host sp$^2$-related defects that act as electron traps and plausible noise sources, reinforcing the view that near-surface disorder and reconstruction can produce persistent electronic states even after standard processing~\cite{stacey2019primalsp2}. 
Taken together, the working consensus is that the surface hosts electronic defect states whose spin and charge dynamics both matter, and whose mitigation requires controlling both chemical termination and near-surface structure.

\paragraph{Measurement artifact: charge conversion can mimic fast $T_1$.}
Near-surface NVs often exhibit charge-state instabilities that directly impact relaxometry readout. Bluvstein \emph{et al.}~showed that shallow NVs can display surface-dependent charge dynamics, including ionization on experimentally relevant timescales, which can degrade ODMR contrast and complicate spin-based measurements~\cite{bluvstein2019charge}. 
For $T_1$ relaxometry in particular, Barbosa \emph{et al.}~quantified how laser-driven conversion between NV$^-$ and NV$^0$ can distort the apparent relaxation: charge conversion occurs even below saturation, and at higher excitation intensities it can dominate the measured fluorescence decay, producing an artificially short $T_1$ if not accounted for~\cite{barbosa2023chargeconversion}. In applied measurements, it is therefore essential to (i) operate at low enough optical power to suppress charge conversion, (ii) normalize fluorescence appropriately, and (iii) whenever possible, utilize differential measurement or monitor NV$^0$ emission in parallel to disentangle spin relaxation from charge dynamics~\cite{barbosa2023chargeconversion,mrozek2015longitudinal,choi2017depolarization,he2023quasi, he2025experimental}.

For nanoscale relaxometry with shallow NVs, surface noise is best treated as an engineered baseline rather than a nuisance. Three practical guidelines recur across the literature:
(i) quantify the baseline by measuring both $T_1$ and $T_2$ (and their power dependence) on the same device and surface condition \cite{rosskopf2014investigation,myers2014probing};
(ii) stabilize the surface by controlling morphology and termination, with ordered oxygen termination as a demonstrated route to substantially improved shallow-NV coherence \cite{sangtawesin2019origins};
(iii) treat charge stability as part of the sensing protocol by choosing excitation conditions and analysis workflows that explicitly separate charge conversion from spin relaxation \cite{bluvstein2019charge,barbosa2023chargeconversion}.
These constraints define the operating envelope in which relaxometry on external targets can be interpreted quantitatively.

\subsubsection{Nanoscale NMR and Nuclear Spin Dynamics}

NV-detected nanoscale NMR turns a near-surface nitrogen-vacancy (NV) center into a local magnetic spectrometer for nuclear spins within a few nanometers of the diamond surface. 
Early ambient-condition demonstrations detected statistically polarized proton ensembles by converting nuclear Larmor precession into an AC magnetic signal and filtering it with dynamical-decoupling sequences~\cite{mamin2013nanoscale,staudacher2013nmr5nm}. 
A key point is that, at the nanoscale, the measured signal is often not the thermal mean magnetization but the root-mean-square spin fluctuations: for $N$ spins, the effective polarization scales as $\sqrt{N}$, so even $N\sim 10^4$ nuclei can produce a detectable fluctuating moment~\cite{meriles2010imaging,herzog2014boundary}. This statistical-polarization regime enables spectroscopy from nanometer-scale detection volumes under room-temperature conditions.

Beyond identifying nuclear species via their Larmor frequencies, NV-based nano-NMR can access \emph{dynamics}. Correlation-type measurements reveal temporal structure in the nuclear field, which encodes molecular motion and local relaxation processes in the near-surface environment \cite{staudacher2015moleculardynamics}. This capability is practically important because nuclear diffusion and surface-driven spectral wandering are often the dominant linewidth mechanisms in liquids near the diamond surface, and they set the performance ceiling for chemical analysis at the nanoscale \cite{staudacher2015moleculardynamics}.

Historically, NV-based NMR spectroscopy has been dominated by dynamical-decoupling readout (e.g., XY8), which interrogates nuclear precession through coherent control and therefore requires microwave instrumentation and nontrivial pulse sequences. A useful counterpoint is that NMR spectra can also be accessed in a purely relaxometric modality, with the NV serving as a field-tunable spectral filter for nuclear spin noise. A particularly instructive example is the microwave-free nano-NMR scheme of Wood \emph{et al.}~\cite{wood2017microwave}. In that work, a shallow NV center is placed under a nanoscale organic layer rich in proton spins, and the bias field is tuned close to the NV ground-state level anti-crossing so that the NV transition frequency enters the MHz regime and becomes resonant with the $^1$H Larmor frequency. Under these conditions, cross-relaxation between the NV and the external nuclear spins enhances the NV longitudinal relaxation rate whenever $\omega_{\mathrm{NV}}(B)$ matches the proton resonance, so that a sweep of $B$ directly yields an NMR spectrum encoded in $T_1(B)$. Crucially, neither the NV nor the nuclear spins are driven by microwaves; the only control is static field tuning and optical initialization/readout. Wood \emph{et al.} showed that this simplified, all-optical protocol achieves a sensitivity comparable to more complex microwave-pulse-based nano-NMR schemes, while probing a $\sim$\,(10\,nm)$^3$ volume of external proton spins under ambient conditions~\cite{wood2017microwave}. This experiment illustrates that NV $T_1$ relaxometry, combined with field-tunable cross-relaxation, can reveal nanoscale NMR spectra in a minimally invasive and experimentally compact fashion. The spectral resolution in $T_1$-based NMR is ultimately limited by the effective linewidth of the resonant transitions, which in many practical cases is set by the dephasing rate $1/T_2^\ast$ (i.e., inhomogeneous broadening). Consequently, the achievable resolution can be poorer than that of coherent, phase-accumulation NV-NMR protocols, where dynamical decoupling enables frequency resolution approaching $1/T_2$ (Fig.~\ref{fig:spin}c-d).

The same principle can, in principle, be extended from point spectroscopy to spatially resolved nuclear-spin imaging: by combining field-tuned $T_1$ relaxometry with either scanning-probe geometry or wide-field NV ensembles, one can construct maps of $T_1(B,\vb{r})$ in which NMR contrast is encoded in the relaxation rate. Such $T_1$-based nano-/micro-NMR complements phase-based NV NMR protocols: relaxometry naturally probes nuclear-spin fluctuations via cross-relaxation, while coherent NV-NMR schemes probe phase accumulation under controlled RF pulse sequences. Together, these techniques provide a unified route to nuclear-spin spectroscopy in which $T_1$ relaxometry supplies a simple, all-optical readout channel with field-tunable spectral selectivity.

In summary, relaxometry-based nuclear-spin spectroscopy with NV centers leverages field-tunable cross-relaxation to encode NMR contrast directly in the longitudinal relaxation rate. By sweeping $B$ to tune $\omega_{\mathrm{NV}}(B)$ across nuclear resonances, one obtains an NMR spectrum as features in $T_1(B)$, without relying on phase-accumulation protocols. Compared with coherent NV-NMR schemes, $T_1$-based NMR is experimentally compact, naturally sensitive to stochastic nuclear fluctuations, and readily compatible with scanning and wide-field implementations. More broadly, relaxometry provides a unified framework spanning spin species and frequency scales, connecting nuclear-spin spectroscopy via cross-relaxation to GHz-band sensing of electronic and magnonic noise in quantum materials.


\section{Conclusion and Outlook}

NV spin relaxometry has evolved from a niche technique into a general-purpose nanoscale noise-spectroscopy platform. Its strength lies in turning complex dynamics into an experimentally accessible decay rate while retaining a quantitative link to the underlying magnetic noise power spectral density. The method is now established across condensed matter (magnons, antiferromagnets, transport noise), spectroscopy (cross-relaxometry and relaxometry-based NMR), and bio/chem sensing (radicals and metal ions), with increasing reach into new host materials and device settings.

For newcomers to the field, the central challenge of spin relaxometry can be viewed as an inverse problem: how can a measured relaxation rate be converted into a microscopic understanding of the underlying dynamics?~\cite{degen2017quantum,casola2018probing,barry2020sensitivity,mzyk2022relaxometry}
In practice, this requires answering several coupled questions. What degrees of freedom generate the magnetic noise? Which part of the noise spectrum is selected by the sensor transition and measurement protocol? How do sensor--sample geometry, near-field propagation, and finite stand-off distance shape the detected signal?~\cite{casola2018probing,kolkowitz2015probing}
How can sample-induced relaxation be distinguished from competing contributions such as surface spins, charge-state fluctuations, laser heating, and technical noise?~\cite{rosskopf2014investigation,myers2014probing,hopper2018spin}
A particularly important practical challenge is that quantitative relaxometry often depends sensitively on the NV--sample distance, which is difficult to determine with high accuracy \cite{xu2025minimizing}.
From a measurement perspective, another limitation is throughput: spin-noise imaging based on $T_1$ relaxometry can be slow, especially when pixel-by-pixel scanning is required.

Addressing these challenges requires advances on multiple fronts. Theoretical and computational efforts are needed to connect measured relaxation rates to microscopic correlation functions, material response functions, and realistic near-field transfer kernels. Materials and surface science will be essential for engineering stable, low-noise sensor interfaces and for reducing spurious relaxation channels. Instrumentation development is required to extend relaxometry to higher frequencies, larger magnetic fields, cryogenic environments, and high-throughput imaging modalities. In parallel, new measurement protocols, including nuclear-ancilla-assisted repetitive readout and spin-to-charge-conversion-based readout, may reduce acquisition time by improving readout fidelity and lowering the number of averages required~\cite{jiang2009repetitive,neumann2010single,holzgrafe2019error,hopper2018amplified}. 
Together, these open problems define a broad research landscape in which physicists, materials scientists, engineers, and data scientists can all make substantial contributions.

Looking forward, a central scientific opportunity is to move from \emph{contrast} to \emph{inference}: from imaging ``where noise is'' to extracting \emph{what noise is} in terms of microscopic mechanisms and material parameters. In practice this means estimating not only a local rate map $\Gamma_1(\mathbf{r})$, but the underlying spectral and spatial structure of fluctuations—e.g., $S_{B_\perp}(\omega)$ and, when relevant, the associated response functions such as $\chi''(\mathbf{q},\omega)$ or $\sigma(\mathbf{q},\omega)$. Achieving this will require measurement sets that deliberately span multiple spectral windows and geometric filters: combining $T_1$, $T_{1\rho}$, and coherence-based probes ($T_2$, dynamical decoupling) across multiple bias fields and multiple sensor--sample distances (depth series, scanning height sweeps, or engineered spacer layers). The payoff is a form of ``noise tomography'' that can disentangle competing channels (surface spins vs.\ sample excitations; magnetic vs.\ electric noise; Johnson noise vs.\ magnetic order-parameter fluctuations) and enable quantitative comparisons to theory and simulation.

A second near-term accelerator is \emph{throughput}. Wide-field cameras already provide parallel relaxometry, but the next step is to make relaxometry fast enough for statistically rich datasets (device variability, spatial heterogeneity, kinetics) and for feedback-based experiments (adaptive field sweeps, real-time mapping during switching, heating, or reaction progress). Promising directions include improved optical readout (spin-to-charge conversion, repetitive readout, and optimized collection optics), lock-in style protocols for suppressing technical drifts, and compressed-sensing / adaptive sampling strategies that target the most informative field points in $\Gamma_1(B)$ rather than uniformly scanning. For scanning geometries, advances in probe stability, drift correction, and real-time height control will be as important as raw sensitivity, because quantitative inversion hinges on reliable knowledge of the sensor--sample distance. For wide-field implementations, recent advances in single-photon avalanche diode (SPAD) array cameras—especially time-gated architectures with nanosecond-scale temporal resolution—can substantially reduce the acquisition time for $T_1$-based imaging \cite{bruschini2024review}.

A third frontier is \emph{integration with extreme or device-relevant environments}. Cryogenic relaxometry can access superconducting quasiparticles, vortex dynamics, and correlated excitations whose noise spectra are sharply temperature and field dependent. At the same time, high-field operation and improved field stability broaden spectral coverage and sharpen cross-relaxation features, enabling more selective spectroscopy (and potentially relaxometry-based microscale NMR in compact geometries). More generally, integrating relaxometry with functioning devices—2D heterostructures under bias, nanoelectronic channels, magnetic tunnel junctions, superconducting circuits, and gated van der Waals magnets—positions relaxometry as a local probe of \emph{nonequilibrium} fluctuations, where the noise itself carries information about dissipation pathways and emergent collective modes.

The \emph{materials landscape} will also broaden. Diamond NV remains the benchmark, but 2D and device-compatible hosts offer complementary advantages: defects in hBN provide intrinsically small standoff and heterostructure compatibility, while SiC offers wafer-scale processing and near-IR emission that is attractive for integrated photonics and bio-adjacent use cases. A practical theme across hosts is that relaxometry performance is often limited by \emph{interfaces} rather than bulk: surface paramagnetism, charge traps, adsorbate dynamics, and electric-field noise can dominate shallow sensors. Thus, progress in surface science (termination control, reconstruction suppression, ultrathin encapsulation layers, and reproducible cleaning/processing) is likely to translate directly into improved relaxometric sensitivity and interpretability.

An additional and increasingly important advantage of spin relaxometry emerges in \emph{high magnetic-field and high-frequency regimes}, where conventional coherent-control techniques become impractical. At bias fields exceeding $\sim$1~T, the NV ESR frequency exceeds $\sim$30~GHz, rendering the delivery of resonant microwave pulses for Rabi driving or dynamical decoupling technically challenging due to the lack of efficient sources, transmission losses, and sample heating. In this regime, standard ODMR- and pulse-based protocols effectively break down. In contrast, all-optical spin relaxometry remains fully operational: the longitudinal relaxation rate continues to sensitively probe magnetic noise from the thermal or driven bath without requiring any resonant RF control, making relaxometry a high–dynamic-range sensing modality that is naturally compatible with extreme fields and frequencies. This capability is particularly relevant for studying high-frequency collective excitations such as antiferromagnetic magnons, whose characteristic frequencies in thin-film systems often span tens to hundreds of gigahertz and are difficult to access using conventional RF-based antiferromagnetic resonance techniques. By applying a strong magnetic field aligned with the sensor’s quantization axis, the ESR frequency of the spin defect can be tuned into resonance with these magnon modes, enabling their detection via cross-relaxation once magnons are excited thermally, electrically, or phononically. More broadly, the ability to operate without applied RF fields makes relaxometry uniquely suited to environments where RF delivery is infeasible or undesirable, including cryogenic platforms, nanoscale devices, and biological or soft-matter systems where RF-induced heating or perturbation must be minimized.

In life-science and chemistry, the most compelling directions go beyond ``detecting radicals'' toward \emph{quantitative, targeted, time-resolved readouts} in complex environments. This includes mapping spatiotemporal radical dynamics during immune response and infection, monitoring redox-active pathways in organelles, and building microfluidic or droplet platforms for higher-throughput assays where relaxometry becomes a screening tool rather than a bespoke measurement. Key enablers will be improved targeting chemistries, calibration strategies that remain valid in heterogeneous ionic environments, and protocols that minimize phototoxicity and local heating. A realistic long-term target is multiplexed relaxometry, where several analytes or microenvironments are discriminated by combining spectral selectivity (field tuning / rotating-frame windows) with engineered spin labels or binding motifs.

Across all applications, several challenges must be addressed for relaxometry to become a broadly quantitative metrology tool. Surface and charge stability remain central: shallow sensors face surface-spin noise and charge conversion that can masquerade as fast $T_1$ or distort apparent contrast, so robust operating envelopes require power-dependent controls, charge-state monitoring, and improved surface preparation/encapsulation. Quantitative inversion and identifiability are equally important: mapping $\Gamma_1(h,B,T)$ back to a unique $S(\omega)$ (or to material parameters) is often ill-posed, and progress will rely on multi-contrast datasets, physically constrained models, and principled uncertainty quantification to avoid over-interpreting non-unique fits. Geometry and distance calibration can be a dominant error source because near-field kernels strongly weight spatial wavelengths and decay rapidly with standoff, so small uncertainties in NV depth or scan height can translate into large errors in inferred parameters. Field, temperature, and drift control are also critical: cross-relaxometry and GSLAC-adjacent nuclear spectroscopy demand stable bias fields and careful management of laser-induced heating and drift, especially in long acquisitions or cryogenic setups. Finally, standardization and benchmarking will be essential—community-wide reference samples, shared analysis pipelines, and standardized reporting (including confidence intervals and control measurements) are needed to compare results across platforms and to build reliable ``noise libraries'' for common materials and surfaces.

In summary, the next phase of NV (and related defect) relaxometry will be defined less by whether a signal can be detected and more by whether it can be \emph{interpreted} and \emph{used}—to extract microscopic dynamics, to diagnose and optimize quantum/mesoscopic devices in situ, and to enable quantitative bio/chemical assays at previously inaccessible length scales. As theory, simulation, and instrumentation co-design mature, relaxometry is poised to become a standard tool for studying dynamical phenomena in quantum materials, nanoelectronics, and living systems.




\section*{Acknowledgments}
This work was supported by the Center for Nanophase Materials Sciences, (CNMS), which is a US Department of Energy, Office of Science User Facility at Oak Ridge National Laboratory. H.Z. and A.M. were supported by the Laboratory Directed Research and Development Program of Oak Ridge National Laboratory, managed by UT-Battelle, LLC, for the U.S. Department of Energy. R.G., G.H., Z.L.~and C.Z.~acknowledge support from the NSF under Grant No. 2514391.

\section*{Author Declarations}
\paragraph{Conflict of Interest:} The authors have no conflicts to disclose.

\paragraph{Author Contributions:} Ruotian Gong, Alex L.~Melendez, Guanghui He, and Zhongyuan Liu contributed equally to this paper.




\bibliographystyle{apsrev4-2}   
\bibliography{references}   

\end{document}